\def\ts     {\thinspace}
\def\kms    {\ifmmode{{\rm \ts km\ts s}^{-1}}\else{\ts km\ts s$^{-1}$}\fi}
\def\msol   {\ifmmode{{\rm M}_{\odot} }\else{M$_{\odot}$}\fi}
\def\lsol   {\ifmmode{L_{\odot}}\else{$L_{\odot}$}\fi}
\def\lfir   {\ifmmode{L_{\rm FIR}}\else{$L_{\rm FIR}$}\fi}

\def\,{\thinspace}
\def \ppm{$\pm$}

\def \ppm{$\pm$}
\def \ly{Ly$\alpha$}
\def \ssa{SSA22}

\documentclass{aastex61}

\usepackage{longtable}
\usepackage{rotating}
\usepackage[switch]{lineno}

\shorttitle{Deep submillimeter and radio observations in the \ssa\, field}
\shortauthors{Ao et al.}

\begin{document}
\title{Deep submillimeter and radio observations in the \ssa\, field. I. 
Powering sources and \ly\, escape fraction of \ly\, Blobs}


\author{Y. Ao}
\affiliation{National Astronomical Observatory of Japan, 2-21-1 Osawa, Mitaka, Tokyo 181-8588, Japan}
\affiliation{Purple Mountain Observatory, Chinese Academy of Sciences, Nanjing 210008, China}

\author{Y. Matsuda}
\affiliation{National Astronomical Observatory of Japan, 2-21-1 Osawa, Mitaka, Tokyo 181-8588, Japan}

\author{C. Henkel}
\affiliation{MPIfR, Auf dem H\"{u}gel 69, 53121 Bonn, Germany}
\affiliation{Astron. Dept., King Abdulaziz Univ., P.O. Box 80203, Jeddah 21589, Saudi Arabia}

\author{D. Iono}
\affiliation{National Astronomical Observatory of Japan, 2-21-1 Osawa, Mitaka, Tokyo 181-8588, Japan}

\author{D. M. Alexander}
\affiliation{Centre for Extragalactic Astronomy, Department of Physics, Durham University, South Road, Durham DH1 3LE, UK}



\author{S. C. Chapman} 
\affiliation{Dept. of Physics and Atmospheric Science, Dalhousie University, Halifax, NS B3H 4R2, Canada}


\author{J. Geach}
\affiliation{Centre for Astrophysics Research, University of Hertfordshire, Hatfield, AL10 9AB, UK}

\author{B. Hatsukade}
\affiliation{Institute of Astronomy, The University of Tokyo, 2-21-1 Osawa, Mitaka, Tokyo 181-0015, Japan}


\author{M. Hayes}
\affiliation{Stockholm University, Department of Astronomy and Oskar Klein Centre for Cosmoparticle Physics, AlbaNova University Centre, SE-10691, Stockholm, Sweden }

\author{N. K. Hine}
\affiliation{Centre for Astrophysics Research, School of Physics, Astronomy and Mathematics, University of Hertfordshire, College Lane, Hatfield, Hertfordshire AL10 9AB, UK}


\author{Y. Kato}
\affiliation{National Astronomical Observatory of Japan, 2-21-1 Osawa, Mitaka, Tokyo 181-8588, Japan}
\affiliation{Department of Astronomy, Graduate school of Science, The University of Tokyo, 7-3-1 Hongo, Bunkyo-ku, Tokyo 133-0033, Japan}

\author{R. Kawabe}
\affiliation{National Astronomical Observatory of Japan, 2-21-1 Osawa, Mitaka, Tokyo 181-8588, Japan}

\author{K. Kohno}
\affiliation{Institute of Astronomy, The University of Tokyo, 2-21-1 Osawa, Mitaka, Tokyo 181-0015, Japan}

\author{M. Kubo}
\affiliation{National Astronomical Observatory of Japan, 2-21-1 Osawa, Mitaka, Tokyo 181-8588, Japan}


\author{M. Lehnert}
\affiliation{Institut d$'$Astrophysique de Paris, CNRS and Universit$\acute{\rm e}$ Pierre et Marie Curie, 98bis Bd Arago, 75014 Paris, France}

\author{M. Malkan}
\affiliation{Department of Physics \& Astronomy, University of California, Los Angeles, CA 90095, USA}

\author{K. M. Menten}
\affiliation{MPIfR, Auf dem H\"{u}gel 69, 53121 Bonn, Germany}


\author{T. Nagao}
\affiliation{Research Center for Space and Cosmic Evolution, Ehime University, Bunkyo-cho 2-5, Matsuyama, Ehime 790-8577, Japan}


\author{R. P. Norris}
\affiliation{CSIRO Australia Telescope National Facility, PO Box 76, Epping, NSW 1710, Australia}
\affiliation{Western Sydney University, Locked Bag 1797, Penrith South, NSW 1797, Australia}




\author{M. Ouchi}
\affiliation{Institute for Cosmic Ray Research, The University of Tokyo, 5-1-5 Kashiwa-no-Ha, Kashiwa City, Chiba 277-8582, Japan}

\author{T. Saito}
\affiliation{Nishi-Harima Astronomical Observatory, Centre for Astronomy, University of Hyogo, 407-2 Nichigaichi, Sayo-cho, Sayo, Hyogo 679-5313, Japan}



\author{Y. Tamura}
\affiliation{Institute of Astronomy, The University of Tokyo, 2-21-1 Osawa, Mitaka, Tokyo 181-0015, Japan}
\affiliation{Department of Physics, School of Science, Nagoya University, Furo-cho, Chikusa-ku, Nagoya, Aichi 464-8602, Japan}

\author{Y. Taniguchi}
\affiliation{The Open University of Japan, 2-11, Wakaba, Mihama-ku, Chiba, 261-8586, Japan}

\author{H. Umehata}
\affiliation{The Open University of Japan, 2-11, Wakaba, Mihama-ku, Chiba, 261-8586, Japan}


\author{A. Weiss}
\affiliation{MPIfR, Auf dem H\"{u}gel 69, 53121 Bonn, Germany}






\begin{abstract}
	We study the heating mechanisms and \ly\, escape fractions of 35 \ly\,
	blobs (LABs) at $z \approx 3.1$ in the \ssa\, field. Dust continuum
	sources have been identified in 11 of the 35 LABs, all with star
	formation rates (SFRs) above 100\msol/yr. Likely radio counterparts are
	detected in 9 out of 29 investigated LABs. The detection of submm dust
	emission is more linked to the physical size of the \ly\, emission than
	to the \ly\, luminosities of the LABs. A radio excess in the
	submm/radio detected LABs is common, hinting at the presence of active
	galactic nuclei.  Most radio sources without X-ray counterparts are
	located at the centers of the LABs. However, all X-ray counterparts
	avoid the central regions. This may be explained by absorption due to
	exceptionally large column densities along the line-of-sight or by LAB
	morphologies, which are highly orientation dependent.  The median \ly\,
	escape fraction is about 3\% among the submm-detected LABs, which is
	lower than a lower limit of 11\% for the submm-undetected LABs.  We
	suspect that the large difference is due to the high dust attenuation
	supported by the large SFRs, the dense large-scale environment as well
	as large uncertainties in the extinction corrections required to apply
	when interpreting optical data.
	
\end{abstract}


\section{Introduction}
\ly\, emission has emerged as a powerful tool to study distant galaxies, as it
is very bright and redshifted to optical wavelengths at high redshifts.  \ly\,
emitters (LAEs), efficiently discovered in narrowband imaging surveys, are
galaxies that emit strong \ly\, radiation presumably from the photoionization
of neutral hydrogen by young, hot stars or active galactic nuclei (AGNs) at
high redshifts. They hold unique clues to the formation and evolution of
galaxies at a time when the Universe was still young (e.g.  Bridge et al.
2013).  However, due to the resonant nature of the \ly\, line and high optical
depth in neutral hydrogen (Hayes 2015), \ly\, photons are likely to undergo
numerous scattering events before they escape from the galaxy or are absorbed
by dust. Thus the actual emitted \ly\, luminosity is a function of the atomic
hydrogen distribution, dust content, gas kinematics and galaxy viewing angle
(Hayes 2015).  Therefore, in order to use \ly\, to study galaxies at high
redshift, we need to understand the escape fraction of \ly\, photons, which is
defined as the ratio of observed to intrinsic \ly\, luminosity and thus
determined by the \ly\, emitter's environment. The evolution of the \ly\,
escape fraction over cosmic time has been determined based on empirical
measurements from large samples (e.g., Gronwall et al. 2007; Ouchi et al. 2008;
Hayes et al.  2011a), providing useful clues to the evolution of the dust
content of galaxies.

As a special class of LAEs, Ly$\alpha$ blobs (LABs) have been most commonly
found in the dense environment of star-forming galaxies at high redshift and
are characterized by their large physical scale (30 to 200 kpc) and high
Ly$\alpha$ luminosity (10$^{43}$ to 10$^{44}$ erg~s$^{-1}$) (see e.g., Francis
et al. 1996; Steidel et al. 2000; Palunas et al. 2004; Matsuda et al.  2004,
2009, 2011, 2012; Dey et al. 2005; Saito et al. 2006; Yang et al. 2009, 2010;
Erb et al.  2011; Prescott et al. 2012a, 2013; Bridge et al. 2013). 
While the LABs' preferential location in overdense environments
indicates an association with massive galaxy formation, the origin of their
Ly$\alpha$ emission is still unclear and under debate (Faucher-Giguere et al.
2010; Cen \& Zheng 2013; Yajima et al. 2013). Proposed sources have generally
fallen into two categories: (1) cooling radiation from cold streams of gas
accreting onto galaxies (e.g., Haiman et al. 2000; Dijkstra \& Loeb 2009;
Faucher-Gigu{\`e}re et al. 2010) and (2) photoionization and/or galactic
super-winds/outflows from starbursts or AGNs (e.g., Taniguchi \& Shioya 2000;
Furlanetto et al. 2005; Wilman et al. 2005; Colbert et al. 2006; Mori \&
Umemura 2006; Matsuda et al.  2007; Zheng et al.  2011; Cen \& Zheng 2013; Ao
et al. 2015; Prescott et al. 2015; Alexander et al.  2016; Hine et al. 2016).
All of the above mentioned energy supplying sources may trigger \ly\, emission
in an environment where violent interactions are frequent between gas rich
galaxies as expected in over-dense regions at high redshift (Matsuda et al.
2009, 2011; Prescott et al. 2012b; Kubo et al.  2013).

Supporting evidence for the cooling flow scenario comes from those LABs lacking
any visible power source (e.g., Smith \& Jarvis 2007).  Dijkstra
\& Loeb (2009) demonstrates that if $>$10 per cent of the change in the gravitational
binding energy of a cold flow goes into heating of the gas then the simulated
cooling flows are spatially extended \ly\, sources that are comparable to
observed LABs. This model can naturally explain the spatial distribution of
the LABs and the diversity of host galaxies in the LABs, as the \ly\, emission
is effectively decoupled from the associated sources. The most luminous
gravitationally powered blobs would be associated with the most massive halos,
which may host a variety of sources like AGNs, Lyman
Break Galaxies (LBGs) and Submillimeter Galaxies (SMGs). 
Alternatively, ionizing photons from young stars in star-forming (SF) galaxies
and/or unobscured AGNs can ionize neutral hydrogen atoms and the subsequent
recombination leads to \ly\, emission. It is usually difficult to discriminate
between the two internal heating mechanisms, SF or AGNs. Resonant scattering of
\ly\, photons in the circumgalactic medium leads to spatially extended emission
(Geach et al.  2005, 2009; Colbert et al. 2006; Webb et al. 2009; Hayes et al.
2011b; Zheng et al. 2011; Cen \& Zheng 2013; Overzier et al. 2013). Cen \&
Zheng (2013) propose an SF-based model and predict that LABs at high redshift
correspond to protoclusters containing the most massive galaxies/halos in the
universe and ubiquitous strong infrared (IR) sources undergoing extreme
starbursts. Their model also predicts that the most luminous FIR source within
each LAB is likely representing the gravitational center of the protocluster.
Note that both cooling flow (Dijkstra \& Loeb 2009) and SF-based models (Cen
\& Zheng 2013) can reproduce the measured luminosity functions of LABs.

To study the heating mechanism(s) of the LABs and investigate their \ly\,
escape fraction, we need to select a large sample of LABs to locate their
accurate positions and investigate their possible powering sources.  \ssa\, is
such a suitable field because it has 35 LABs detected at $z \approx 3.1$
(Matsuda et al. 2004), providing an ideal laboratory to study the LABs in a
large sample. One of them, SSA22-LAB01 is the best studied source (e.g.,
Matsuda et al. 2007; Yang et al. 2012; Geach et al. 2014, 2016; Umehata et al.
2017a). At (sub)mm wavelength, most LABs of \ssa\, have been studied with the
Submillimetre Common-User Bolometer Array (SCUBA, Geach et al. 2005) and
SCUBA-2 (Hine et al. 2016) on the James Clerk Maxwell Telescope (JCMT), and
with the AzTEC 1.1~mm camera (Tamura et al. 2009, 2013; Umehata et al. 2014) on
the Atacama Submillimeter Telescope Experiment (ASTE). However, no
significant 1.1~mm continuum has been found in any of individual 35 LABs
(Tamura et al.  2013).  Even in the recent deep SCUBA-2 observations, only two
out of 34 LABs are detected at 850~$\mu$m (Hine et al.  2016). A few LABs have
been observed and detected in the dust continuum with the Atacama Large
Millimeter/Submillimeter Array (ALMA; Umehata et al.  2015; Alexander et al.
2016; Geach et al. 2016).  In this paper, we present deep submm results from
new JCMT/SCUBA-2 data (Holland et al. 2013) together with ALMA data, and deep
radio images from the Karl G. Jansky Very Large Array (VLA)\footnote{The
National Radio Astronomy Observatory is a facility of the National Science
Foundation operated under cooperative agreement by Associated Universities,
Inc.} observations to study the LABs in the \ssa\, field. Note that in this
paper we only focus on the LABs; the SMGs in this region will be presented in
an upcoming paper based on the same submm and radio data (Ao et al. in
preparation).

\section{Observations}
\subsection{JCMT/SCUBA-2 observations}
The observations were carried out at 850~$\mu$m with SCUBA-2 (Holland et al.
2013) at the JCMT. The data were taken between April 20 and June 30 in 2015
under good weather conditions when the zenith optical depth at 225~GHz was in
the range $0.04<\tau_{\rm 225}<0.08$ with a mean $<\tau_{\rm 225}>$ of 0.06. We
observed a subregion of \ssa\, with a total on-source observing time of 19
hours, covering a field with a diameter of 15 arcmin, centered on $\alpha(\rm
J2000)$\,=\,22$^{\rm h}$17$^{\rm m}$31$^{\rm s}$.7, $\delta(\rm
J2000)$\,=\,+00$^{\rm o}$17$\arcmin$50$\arcsec$, using multiple repeats (40
mins per repeat) of the PONG scanning pattern (Holland et al. 2013).  The \ssa\,
field was also observed as part of the JCMT SCUBA-2 Cosmology Legacy Survey
(S2CLS, Hine et al.  2016; Geach et al. 2017) with a total on-source observing
time of 72 hours to cover a map with a diameter of 30 arcmin, centered on
$\alpha(\rm J2000)$\,=\,22$^{\rm h}$17$^{\rm m}$36$^{\rm s}$.3, $\delta(\rm
J2000)$\,=\,+00$^{\rm o}$19$\arcmin$22$\arcsec$.7. The observations are
summarized in Table~\ref{table_log}.

Pointing checks and flux calibration were achieved via observations of Neptune
and Uranus, immediately before and after the science exposures. Data reduction
was carried out using the SubMillimeter User Reduction Facility (smurf) {\it
makemap} pipeline (Chapin et al. 2013), with flat-fields, image stacking, and
removing atmospheric emission (see also Hine et al. 2016 for more details).
The main beam size of the SCUBA-2 observations at 850~$\mu$m is 14$\arcsec$ and the
map is convolved with a smoothed beam of 30$\arcsec$ to optimize the
detection of point sources.  The total on-source integration time was about 91
hours (see Table~\ref{table_log}). The final beam-convolved map reaches an rms
noise level of 0.75~mJy/beam within the central 15 arcmin and about
1.0~mJy/beam outside the central region (see Figure~\ref{scuba_rms}).

\subsection{ALMA observations}
ALMA observations were carried out in band 7 with a central frequency of 350
GHz towards 4 LABs (LAB1, LAB2, LAB5 and LAB18) (project code: 2013.1.00704S;
Matsuda et al. in preparation).  Another two LABs, LAB12 and LAB14, had been
covered by a deep field in \ssa\, (Umehata et al. 2015, 2017b) in band 6 with a
central frequency of 263 GHz. The data were reduced with the CASA package in a
standard manner (for the details see Umehata et al. 2017b and Matsuda et al. in
preparation).

\subsection{VLA observations}
We observed the \ssa\, region centered around the location of our SCUBA-2
observations with the VLA in B-configuration at S-band (2-4 GHz), under
projects 15A-120 and 16A-310. During 17 sessions with a total observing time of
41 hours, we observed three positions in 2015 and 2016 (see
Table~\ref{table_log} for the observing log).  The total bandwidth was 2 GHz,
split into 16 spectral windows.

\begin{figure*}[t]
\centering
\includegraphics[angle=0,width=1.0\textwidth]{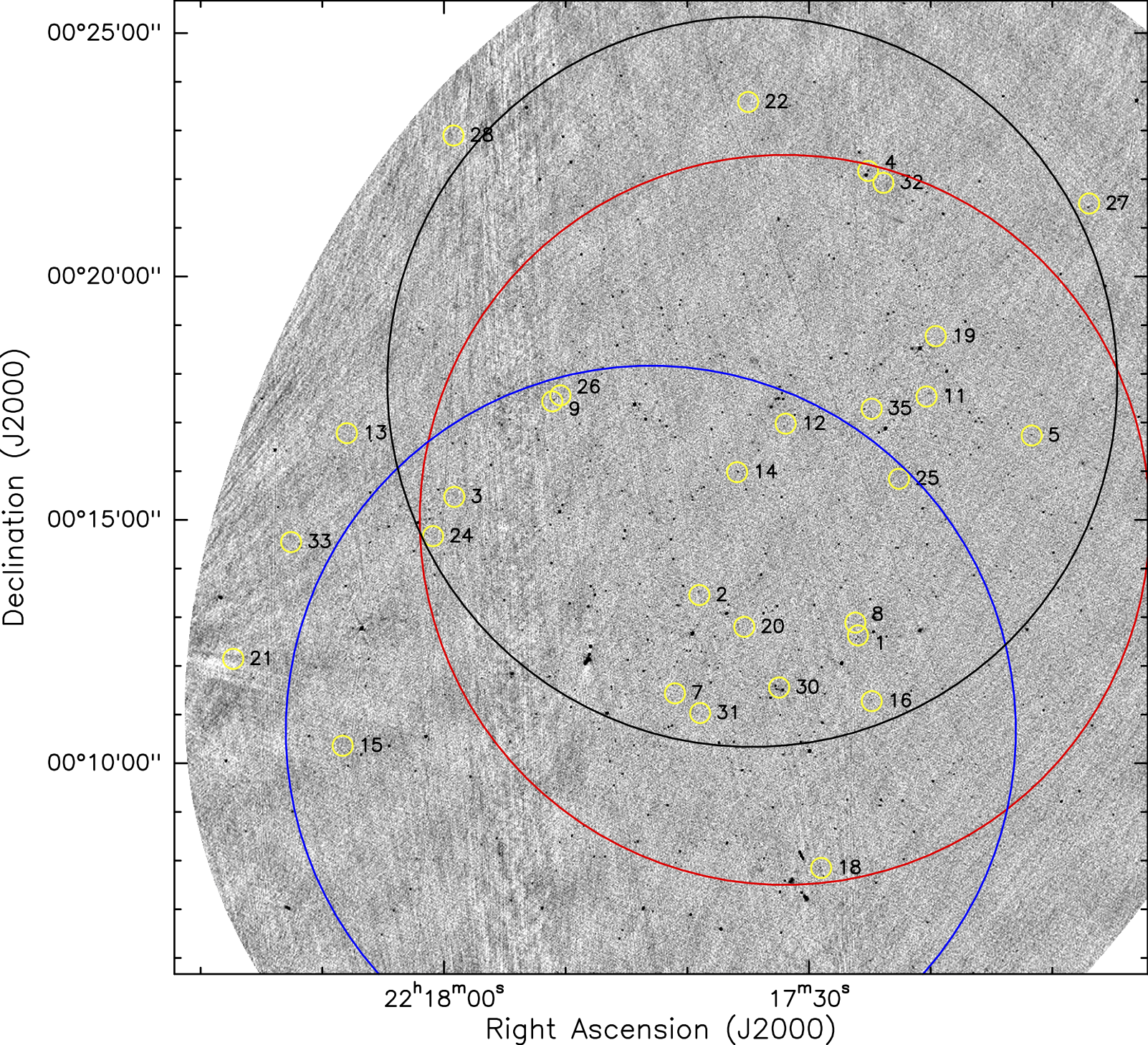}
\caption{A radio map of the \ssa\, region in grey scale. Three big circles
	denote the primary full width to half maximum beam size for three
	pointings with different on-source observing time (red: 26 hours, blue:
	1.6 hours and black: 4.2 hours). The locations of 29 out of the 35 LABs
	identified by Matsuda et al.  (2004) are covered by our observations
	and are shown as yellow circles with diameters of 25
	arcsec. The ID numbers of the sources are labelled in
	black.}\label{radio_rms}
\end{figure*}

Data were first processed through the VLA Common Astronomy Software
Applications (CASA) Calibration Pipeline by NRAO staff, performing basic
flagging and calibration. Then we iteratively inspected the data, and then
flagged the data with radio frequency interference. The final mosaic of images
was created with CASA task CLEAN, reaching an rms sensitivity of
1.5~$\mu$Jy/beam (before primary beam correction) and an angular resolution of
2.3$\arcsec\times$2.0$\arcsec$. The primary beam of one single pointing is
about 15 arcmins. In Figure~\ref{radio_rms}, we indicate the locations of 
29 out of the 35 LABs identified by Matsuda et al. (2004) in the final radio image.

\section{Results}

\subsection{Submm emission}
\begin{figure*}[t]
\centering
\includegraphics[angle=0,width=1.0\textwidth]{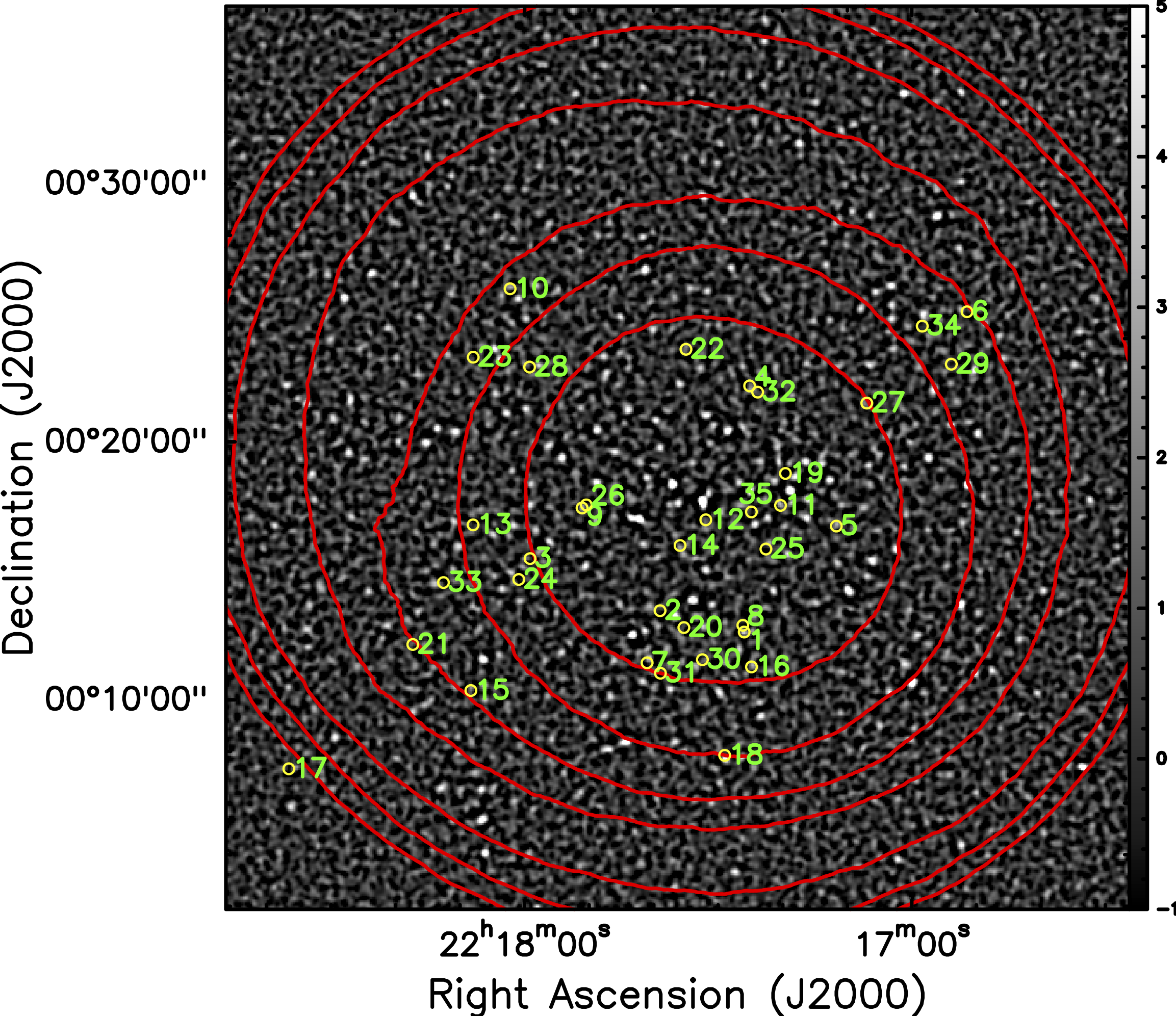}
	\caption{The signal-to-noise map of the SCUBA-2 data in the \ssa\,
	region in grey is overlaied with contour levels of 0.8, 0.9,
	1.0, 1.1, 1.3, 1.5 and 1.7 mJy/beam of the noise map. The locations of
	the 35 LABs identified by Matsuda et al. (2004) are shown as yellow
	circles with diameters of 25 arcsec. The ID numbers of the sources are
	indicated in green.}
\label{scuba_rms} 
\end{figure*}

For the 35 LABs identified in Matsuda et al. (2004), all sources are covered by
SCUBA-2 observations, as shown in Figure~\ref{scuba_rms}.  The \ssa\, field has
been observed as part of the JCMT S2CLS project (Geach et al. 2017), reaching a
1$\sigma$ level of 1.1 mJy/beam at 850~$\mu$m. Combined
with our SCUBA2 observations in 2015, we reach a deeper rms sensitivity of
about 0.75 mJy/beam for the overlapping region.

The SCUBA-2 850~$\mu$m flux measurements are listed in Table~\ref{table_obs}.
9 out of the 35 LABs are detected in dust emission with peak SNRs above
2.5$\sigma$, and 6 of them above 3$\sigma$ (see Figure~\ref{scuba_radio}). The
three sources, LAB4, LAB9 and LAB14, with only marginal detection levels of
2.5$\sigma$ have all been detected in the radio above 5$\sigma$. 

Figure~\ref{scuba_alma} shows the SCUBA-2, VLA and ALMA images together to
cross check the detections with low significance from the SCUBA-2 and VLA
observations by the highly significant ones from ALMA. The typical pointing
accuracy of the JCMT is about 1-2$\arcsec$.  Due to the noise of the image,
the positional uncertainty of the SCUBA-2 observations is related to its
signal-to-noise ratio (SNR) and beam size (Condon et al. 1998) via $\sigma_{\rm
p}=\frac{\theta_{\rm beam}}{{\rm SNR}\,(2{\rm ln}2)^{0.5}}$, where $\sigma_{\rm
p}$ is the 1-$\sigma$ positional uncertainty and $\theta_{\rm beam}$ the beam
size.  For a marginally detected source with an SNR of 3 and a smoothed beam of
30$\arcsec$, its positional uncertainty will be about 8.5$\arcsec$.  Six LABs
observed by ALMA have been detected with high SNRs. Two sources, LAB2 and
LAB12, have not been detected with SCUBA-2 due to low flux densities, but have
been detected with ALMA in the dust continuum. For the remaining four LABs,
their positions are consistent with our SCUBA-2 observations if considering
the positional uncertainties of SCUBA-2 observations. The VLA positions of the
radio emission are in good agreement with those of the dust emission obtained
with ALMA. These consistent results support the reliability of the SCUBA-2 and
VLA data, even for the marginal detections. Adopting the number
count study of SCUBA-2 sources in Geach et al. (2017), the probability of
finding a 850~$\mu$m source with a flux greater than 2 mJy within a box of
15$\arcsec$, accounting for the typical LAB's size and the SCUBA-2's beam size at
850$\mu$m, is $\sim$5.2\%.  Thus, we expect to have two spurious submm sources
detected above 2mJy among the 35 LABs in the \ssa\, field. However, all
SCUBA-2-detected sources have significant radio counterparts except for LAB10
and LAB11. The former is not covered by the VLA observations and the latter has
a weak radio counterpart. Thus, all SCUBA-2 sources are very likely associated
with the LABs, instead of merely representing chance coincidences along the
given lines-of-sight.

Previously, three deep (sub)mm surveys had been carried out in
this field (Geach et al. 2005, 2017; Tamura et al. 2013; Hine et al. 2016) and
their results are presented in Table~\ref{table_obs}. Early SCUBA data (Geach
et al.  2005) show five sources detected at $\ge$3.5$\sigma$. However, only two
sources, LAB1 and LAB18, have been confirmed by the recent deeper SCUBA2
observations (Hine et al. 2016). Our results are consistent with those in Hine
et al. (2016). For the LABs detected in Geach et al. (2005), we confirmed all
sources but mostly with much lower flux densities. The large difference may be
due to flux boosting in the original SCUBA data (Chapman et al. 2001) and
issues related to data reduction and calibrations (Hine et al. 2016). Adopting
the SED templates described in $\S$~\ref{sfr}, one expects that flux densities
at 1.1~mm are about half of those at 850~$\mu$m. Thus, it is not
surprising that none of the LABs is individually detected at $\ge$3.5$\sigma$
at 1.1~mm by AzTEC/ASTE (Tamura et al. 2013). Typical predicted fluxes of
SCUBA2-detected LABs are less than 1.5 mJy at 1.1~mm, which corresponds to
about 2$\sigma$.  Actually, the brightest source, LAB18, is the only 3$\sigma$
detection at 1.1~mm, and is consistent with the predicted value from the SED
templates.

We also stack the remaining submm-undetected LABs except for
LAB17, because the latter is located in a region with a high noise level (see
Figure~\ref{scuba_rms}). The stacked image shows no significant detection at
$>$3$\sigma$ ($~$0.56~mJy).

\subsection{Radio emission}
In Figure~\ref{radio_rms}, we show the full radio map imaged by the VLA
observations with the radio emission in grey scale.  29 sources are covered by
our observations, and 7 (LAB13, LAB15, LAB21, LAB22, LAB27, LAB28 and LAB33) of
them are outside the central Field of View (FoV) with a noise level of above
$\sim$3.5~$\mu$Jy/beam after primary beam correction. None of the sources
outside of the FoV are detected at radio wavelengths, and this may well be related to
the lower sensitivity in these regions. Among the remaining 22 LABs, 9 are
detected above 4$\sigma$. The VLA S band flux measurements are listed in
Table~\ref{table_obs}. For 5 out of 9 radio detected sources, spectroscopic
data from the literature at the same locations as the radio counterparts 
show that their redshifts are around 3.1 (see Table~\ref{table_obs}). Adopting
the number count study of radio sources in Condon et al. (2012), the
probability of finding a 3 GHz source with a flux greater than 7.5 $\mu$Jy
within a typical LAB's size of 6$\arcsec$ is $\sim$4.2\%. Thus, among the 22
sources with good sensitivities, we expect to have one spurious radio source
detected above 7.5 $\mu$Jy.

\subsection{Star formation rates}\label{sfr}
Here we derive the star formation rates (SFRs) from the \ly, IR and radio
luminosities. To estimate the SFR from the \ly\, luminosity, we first assume
that star formation (SF) powers the observed \ly\, flux.  We use an unreddened
Ly$\alpha$/H$\alpha$ ratio of 8.7:1 and the conversion factor between H$\alpha$
luminosity and SFR (Kennicutt 1998; Kennicutt \& Evans 2012), yielding
SFR($\rm{Ly\alpha}$)/(\msol/yr)\,=\,$0.62~\times\,L_{\rm Ly\alpha}$/(10$^{42}$
erg s$^{-1}$). This provides a lower limit, because the dust extinction of \ly\,
emission, likely exacerbated by resonance scattering, may significantly reduce
the observed \ly\, luminosity.

For another estimate of the SFR, we first need to determine the IR luminosity.
However, the dust SEDs of the LABs cannot be well constrained, as only one or
two measurements at (sub)millimeter wavelengths are available.  Therefore, we
will follow the method described in Umehata et al. (2015) to use SED templates
of well studied starburst galaxies, Arp 220 and M82 (Silva et al.  1998), a
composite SED of SMGs from the ALMA LESS survey (ALESS, Swinbank et al. 2014),
and SMM J2135$–$0201 (the cosmic eyelash; Swinbank et al. 2010) to consider a
variety of SEDs. We created best fit SEDs for each template based on redshift
and SCUBA-2/ALMA measurements. The spectra between 8 and 1000~$\mu$m in the
rest frame were integrated, and we derive a median value as well as
minimum/maximum values. Following the star formation rate calibration in
Kennicutt (1998) and Kennicutt \& Evans (2012), we can estimate the SFR by
using the relation SFR($L_{\rm FIR}$)/(\msol/yr)\,=\,$1.46\times$$L_{\rm
FIR}$/(10$^{10}$ \lsol).  The uncertainties of submm-derived SFRs mainly come
from the choice of adopted templates, and the minimum and maximum as well as
median values are given in Table~\ref{table_sfr}.

In luminous galaxies, radio emission is dominated by synchrotron radiation from
electrons, and one can relate this emission to the SFR by SFR($L_{\rm
1.4~GHz}$)/(\msol/yr) = 5.52$\times$10$^{-22}$~$L_{\rm 1.4~GHz}$/(W Hz$^{-1}$)
(Bell 2003). The radio luminosity at 1.4 GHz in the rest frame can be estimated
from the observed flux at 3 GHz by assuming a relation $S \propto \nu^\alpha$,
where S is the flux density and a typical spectral index commonly adopted for
SMGs (e.g., Ivison et al.  2010) is $\alpha$ = $-$0.8. The star formation rates
derived using these three methods are listed in Table~\ref{table_sfr}.

\subsection{Comments on individual LABs}\label{individual}
Radio data are not only helpful to discover the powering sources, but also to
provide accurate positions of the unresolved SCUBA-2-detected
sources and help to cross-identify and even verify the corresponding dust
emission at a relatively faint detection level. Here, we will briefly describe
the sources detected at submm and radio wavelengths. Deep X-ray observations
with $Chandra$ are available for the SSA22 field (Geach et al. 2009; Lehmer et
al. 2009), and will also be discussed for the detected sources. For
those readers mainly interested in statistically relevant results, we recommend
continuing with $\S$~\ref{discussion}.

\subsubsection{LAB1}
For LAB1, the ALMA observations at 850~$\mu$m reveal three cores with a total
flux density of 1.72\ppm0.21 mJy (upper left panel in Figure~\ref{scuba_alma};
see also Geach et al. 2016 and Matsuda et al. 2017), two close to VLA-LAB1a and
one close to VLA-LAB1b. Using the SCUBA-2 data, Geach et al. (2014) found a
flux density of 4.6\ppm1.1 mJy.  Not accounting for the slightly different
SCUBA-2 and ALMA primary beams, this suggests that 63\ppm10\% of the extended
emission is missed by the ALMA observations. Combining this with our new SCUBA-2
data, we find a flux density of 2.9\ppm0.8 mJy, indicating that missing flux
accounts for 41\ppm15\% of the total flux density. Considering the flux
uncertainties, the new value is only slightly higher than ALMA's measurement,
showing that in this LAB there may be not much extended structure missed by
ALMA. Adopting the SED templates described in $\S$~\ref{sfr}, the
predicted flux densities are 0.46$^{\rm +0.40}_{\rm -0.12}$ mJy for ALMA-LAB1ab,
0.27$^{\rm +0.24}_{\rm -0.07}$ mJy for ALMA-LAB1c at 1.25~mm, and
0.025$^{\rm +0.022}_{\rm -0.007}$ mJy for ALMA-LAB1 at 3.5~mm, respectively.
These results are consistent with the 3$\sigma$ upper limits around LAB1 of
0.45~mJy at 1.25~mm and 0.15~mJy at 3.5~mm reported by Yang et al. (2012).

Two radio sources are detected in this LAB. The northern radio source,
VLA-LAB1a, peaks at a location close to two dust continuum peaks, ALMA-LAB1a
and ALMA-LAB1b. The southern radio source, VLA-LAB1b, is consistent with one of
the ALMA 350 GHz continuum sources, ALMA-LAB1c. Weak X-ray emission is detected
around the southern source. [C{\scriptsize II}] emission  has also been
detected in ALMA-LAB1b with a secured redshift of 3.0993\ppm0.0004 (Umehata et
al.  2017b).

\subsubsection{LAB2}
Radio emission is detected within this LAB. However, SCUBA-2 observations show
no dust emission around this source down to a 2$\sigma$ level of 1.6 mJy.
Sensitive ALMA observations find a counterpart near the radio source with a
flux density of 0.91\ppm0.10 mJy at 350~GHz, coincident with the X-ray
counterpart (Geach et al. 2009). 

Our observation does not confirm the presence of a second continuum source (see
Figure~\ref{scuba_alma}) marginally detected by Alexander et al. (2016) with a
flux of 1.11\ppm0.25 mJy at 0.87mm, which is significantly higher than our 3$\sigma$
ALMA limit of 0.23 mJy. This may be due to the presence of extended structure
that is resolved out by our higher angular resolution observations or it may be
a spurious source.

\subsubsection{LAB3}
The VLA data show a tentative detection within this LAB, but no dust emission
is detected with SCUBA-2. Note that the radio peak has only a 3$\sigma$
significance, and there are no counterparts at other wavelengths. Thus, the VLA
signal might not be reliable, and we consider this LAB as undetected at radio
frequencies. A strong X-ray source is detected in this LAB, however it is not
coincident with the weak radio peak.

Recently, dust continuum has been marginally detected around this LAB with ALMA
(Alexander et al. 2016), but it is outside of the LAB, offset $\sim$4.5 arcsec
($\sim$30~kpc) from the center. Thus, the dust continuum is not considered to
be associated with the LAB.

\subsubsection{LAB4}
The radio emission peaks at the center of the LAB. However, the SCUBA-2 submm
source is offset 9 arcsec from the center. Considering the positional
uncertainty of about 9 arcsec for this weak source, we still consider the submm
source to be consistent with the radio source and also associated with the LAB.
A similar situation is encountered in LAB14, where the ALMA observations
confirm the association between the radio and submm sources. Note
that there is a bright radio source at the southeastern edge
(Figure~\ref{scuba_radio}). This is possibly a foreground source.

\subsubsection{LAB5}
This LAB is detected by SCUBA-2 and the VLA, and further confirmed by the ALMA
observations.  The SCUBA-2 and ALMA data show similar flux densities for this LAB.

\subsubsection{LAB9}
This LAB shows dust emission around the center. The VLA data reveal two radio
components within this source, but only one source with an SNR above
4$\sigma$.

\subsubsection{LAB10}
This source is detected at submm wavelengths with a significance of 3$\sigma$
by SCUBA-2, but is not covered by our radio observations. Further
cross-identification is needed to confirm its reliability.

\subsubsection{LAB11}
This LAB shows SCUBA-2 dust emission around the center. The marginally detected
radio emission shows elongated structure, and its reality needs to be confirmed
because of its low SNR.

\subsubsection{LAB12}
A significant radio detection is found within the LAB, but no dust counterpart
is detected by SCUBA-2. However, the more sensitive ALMA observations at
1.14~mm show dust emission around the radio source. It also coincides with a
strong X-ray counterpart (Geach et al. 2009). This LAB is also
detected at 0.87~mm with ALMA by Alexander et al. (2016), showing a flux density of
1.58\ppm0.35~mJy. Both ALMA measurements can be well fitted by the SED templates
described in $\S$~\ref{sfr}.

\subsubsection{LAB14}
The radio emission peaks around the center of the LAB, and SCUBA-2 shows dust
emission centered 7 arcsec off the radio peak. However, our ALMA observations
demonstrate that the 1.14~mm dust emission coincides with the radio counterpart
(see Figure~\ref{scuba_alma}), supporting the reality of the SCUBA-2
detection.  The inconsistency between the dust seen by SCUBA-2 and ALMA as well
as the radio emission may be explained by the 1-$\sigma$ positional uncertainty
of 9$\arcsec$ of the SCUBA-2 data. The radio source is also coincident with a strong
X-ray counterpart (Geach et al.  2009).

This LAB was also detected at 0.87~mm by ALMA (Alexander et al.
2016), showing a flux density of 2.96\ppm0.29~mJy. The SED templates described
in $\S$~\ref{sfr} can well fit both ALMA measurements. It predicts a flux
density of 3.12\ppm0.31 at 0.85~mm, which is about 56\% higher than the SCUBA-2
measurement, suggesting that SCUBA-2 only reveals about two third of the total
flux. It may be due to the effect of a negative sidelobe from a bright source
in the south, which is 16$\arcsec$, close to the SCUBA-2 beam size of
14$\arcsec$, off the SCUBA-2 source associated with LAB18.

\subsubsection{LAB16}
The radio emission is detected around the center of the LAB and SCUBA-2 shows
dust emission that peaks around 5 arcsec from the center, which is well within
the 1-$\sigma$ positional uncertainty of the SCUBA-2 data.

\subsubsection{LAB18}
This is the strongest submm source among all LABs in this field. The SCUBA-2
observations reveal a flux density of 5.4\ppm0.9 mJy. The ALMA observations
discover four dust cores (a, b, c, and d in the lower right panel of
Figure~\ref{scuba_alma}) with a total flux density of 9.47\ppm0.36 mJy, which
is about 75\% higher than the SCUBA-2 value.  Three out of four ALMA submm
sources have radio counterparts. ALMA-LAB18a peaks around the center of the LAB
and ALMA-LAB18b lies south of the center with an offset of 4.5 arcsec. The
latter is surrounded by \ly\, emission. If the \ly\, emission in the south is a
part of LAB18, its elongation along the north-south direction is around 13
arcsec, i.e., 100 kpc.  The other two submm sources, ALMA-LAB18c and
ALMA-LAB18d, located farther to the south are outside the LAB and not
associated with it. Geach et al. (2009) found an X-ray counterpart between two
ALMA sources, ALMA-LAB18b and ALMA-LAB18c. However, the positional errors of
the X-ray sources are expected to be the order of 2.5 arcsec in most cases
(Lehmer et al.  2009). However, LAB18 is at the edge of the X-ray image.
Therefore, its positional error will be even larger, and the X-ray source might
be associated with either ALMA-LAB18b or ALMA-LAB18c.

There is no continuum emission detected at 3.55~mm by Yang et al.
(2014), showing a 3$\sigma$ upper limit of 0.13~mJy. This is confirmed by the
predicted flux density of 0.054$^{\rm +0.048}_{\rm -0.014}$ mJy from the SED
templates.

\subsubsection{LAB30}
There are four radio sources detected in this region
(Figure~\ref{scuba_radio}). The sources, A, C and D, show detections of high
significance but are located outside of this LAB.  Source D is found to be
associated with a local galaxy at z=0.41 (Saez et al. 2015).  Source B is
close to the center of the LAB, but only marginally detected at 3$\sigma$.  No
submm emission is detected and the marginal detection of source B needs to be
confirmed. We consider this source as undetected at radio wavelengths.

\section{Discussion}\label{discussion}

\subsection{Physical sizes of LABs and detection of submillimeter emission}

\begin{figure*}[t]
\centering
\includegraphics[angle=0,width=1.0\textwidth]{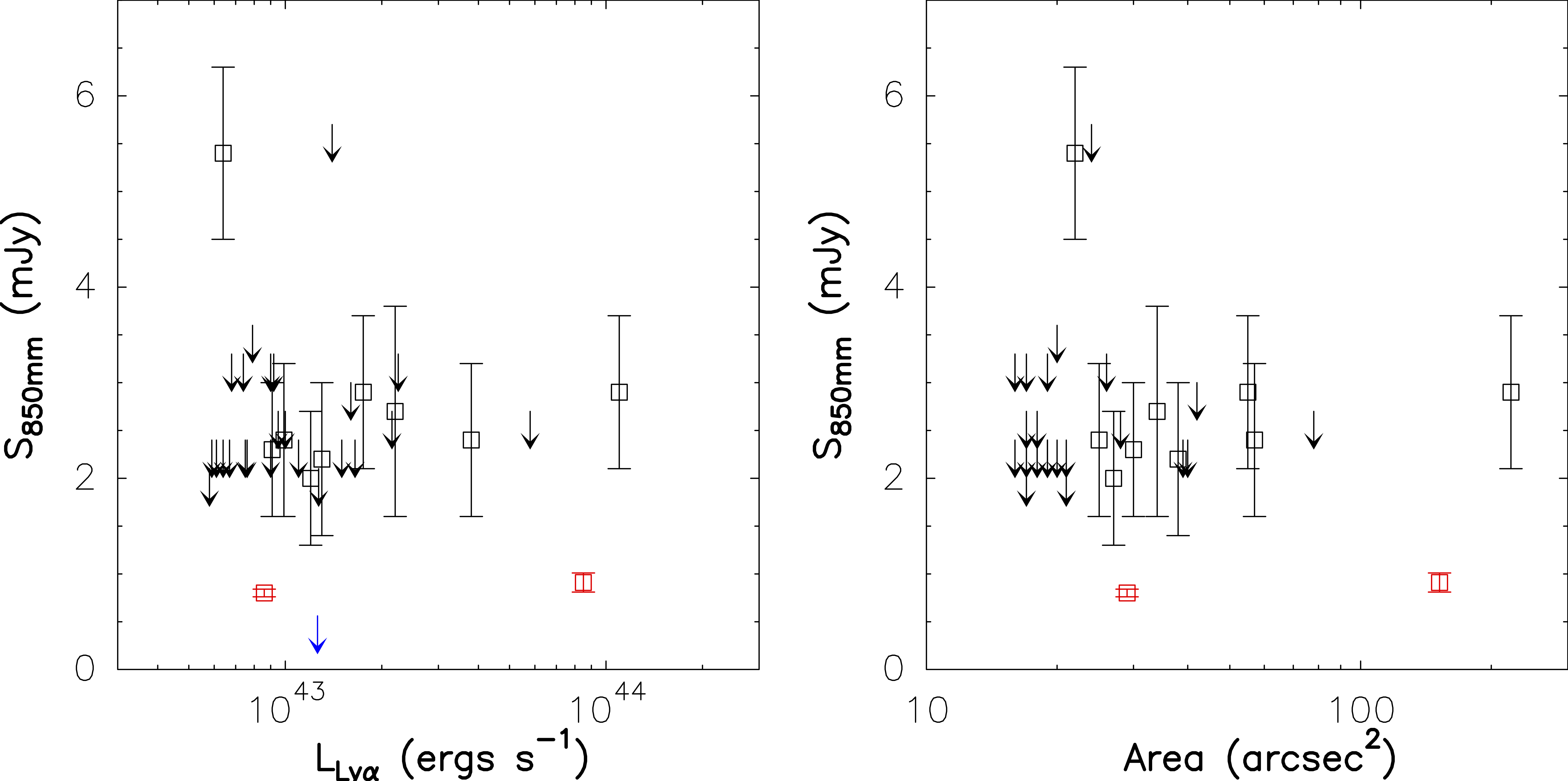}
	\caption{Distribution of measured 850$\mu$m flux versus \ly\,
	luminosity ({\bf left}) and isophotal area ({\bf right}) for the LABs
	in the \ssa\, field. The black empty squares denote the LABs detected at
	850$\mu$m and the arrows mark undetected sources with 3$\sigma$ upper
	limits. Two sources, LAB2 and LAB12, are only detected by ALMA with
	850$\mu$m fluxes less than 1.0 mJy, shown as red squares.
	}\label{size} 
\end{figure*}

Figure~\ref{size} shows the distribution of \ly\, luminosity, isophotal area
and measured fluxes at submm wavelengths for all LABs. In all 11
submm-detected sources, two ALMA-detected LABs, LAB2 and LAB12, remain
undetectable by our SCUBA-2 observations, while the remaining 9 sources are
detected by SCUBA-2.  In this section, we will therefore rely on the SCUBA-2
data alone to investigate possible relations between the physical size of LABs
and the detection of submm emission. In the left panel, we plot the submm flux
density against the \ly\, luminosity. For the sources with $L_{\rm
Ly\alpha}$$>$10$^{43}$\,ergs~s$^{-1}$, 6 out of 17 LABs are detected at submm
wavelengths with SCUBA-2. For the fainter sources with $L_{\rm
Ly\alpha}$$<$10$^{43}$\,ergs~s$^{-1}$, 3 out of 18 LABs are detected at submm
wavelengths. Apparently, more \ly\, luminous galaxies are more likely to be
detected at submm wavelengths. In the right panel, we plot the submm flux
density against the isophotal area.  For the first 18 large LABs, which are
named according to their isophotal area in Matsuda et al. (2004), 9 are
detected in the submm. The remaining 17 LABs are not detected at submm
wavelengths. This suggests that large LABs exhibit stronger submm emission.
Furthermore, our results suggest that the detection of submm dust emission in
LABs is more linked to the physical sizes of \ly\, emission than to their \ly\,
luminosities. The physical sizes of \ly\, emission, instead of
their luminosities, may have a tight correlation with dust emission. This seems
to be inconsistent with the fairly tight correlation between the physical sizes
of LABs and the \ly\, luminosities reported by Matsuda et al.  (2004). However,
one should note that the correlation is weak if excluding the four
brightest/largest LABs. The (sub)mm emission can be a good indicator of gas
masses for SMGs because of its low optical depth.  Due to the uncertainties in
the \ly\, escape fraction of LABs, the \ly\, emission is not a good mass tracer
of SMGs and therefore the correlation between the dust emission and the \ly\,
emission is not very tight.  The larger physical extension of \ly\, emission
may be related to spatially more extended surrounding atomic hydrogen and star
forming regions emitting ionizing photons. A recent study (Matsuda et al.
2012) shows that the spatial extent of the \ly\, haloes is determined by the
surrounding Mpc-scale environment, rather than by the central UV luminosities.
However, this is inconsistent with the result in Xue et al. (2017), who don't
find any correlation between measured scale-lengths and degree of overdensity
relative to the environment. It is worth to emphasize that the
trend to dust detections, more related to LAB sizes than to LAB luminosities,
is based not entirely on detections but on detections of the large area LABs
and non-detection of the smaller area LABs. Future more sensitive ALMA data may
settle this puzzle and reveal the relation between LAB sizes and their possible
internal heating sources in a more convincing way.

\subsection{Powering sources of LABs: star formation or active galactic nuclei?}

\subsubsection{Submm and radio detections in LABs and their implications to
heating mechanisms}\label{submm_radio_discussion}
The heating mechanism of LABs is still unclear. About one third of the LABs in
\ssa\, are associated with submm/radio sources with SFRs above 100~\msol/yr.
Spectroscopic measurements from the literature (see Table~\ref{table_obs})
confirm that at least some of the detected submm/radio counterparts are indeed
associated with the LABs. These results suggest that internal heating in the
host galaxies may be a major energy source for some LABs. It is consistent
with our recent study of J2143$-$4423 (Ao et al. 2015), where 2
out of 4 LABs have been detected at radio and submm wavelengths.

\begin{figure*}[t]
\centering
\includegraphics[angle=0,width=1.0\textwidth]{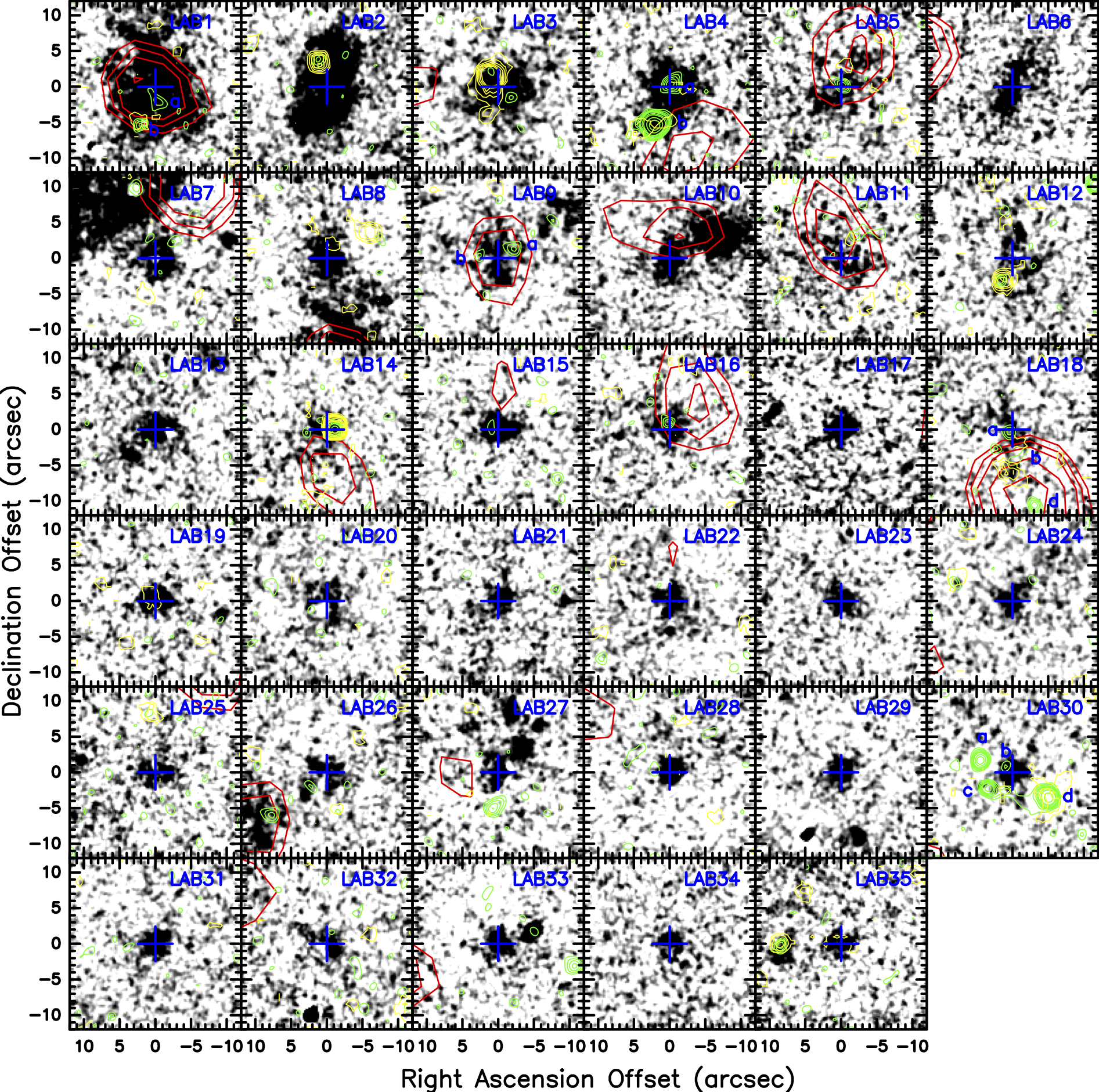}
	\caption{SCUBA2 dust emission at 850~$\mu$m in red (the contour levels
	are 2, 2.5, 3, 4 and 5 times the respective noise level, the latter
	given in Column~4 of Table~2), VLA radio emission at 3~GHz in green
	(contour levels are 2, 3, 4, 5, 6, 8, 10, 15, 20 and 30$\sigma$ and the
	values of $\sigma$ are the noise levels prior to primary beam
	correction given in Column~6 of Table~2) and $Chandra$ X-ray emission
	in yellow (the contour levels of number counts of the smoothed
	full-band, 0.5-8~keV, image adopted from Lehmer et al.  2009 are 6, 9,
	12, 15, 18, 21, 30, 40 and 50) are overlaid on the \ly\, emission taken
	from Matsuda et al. (2004).  Radio sources are labelled with letters in
	blue when more than one radio core is detected in one LAB.  The
	offsets are relative to the centers of the LABs, denoted by blue
	crosses.}\label{scuba_radio} 
\end{figure*}

\begin{figure*}[t]
\centering
\includegraphics[angle=0,width=1.0\textwidth]{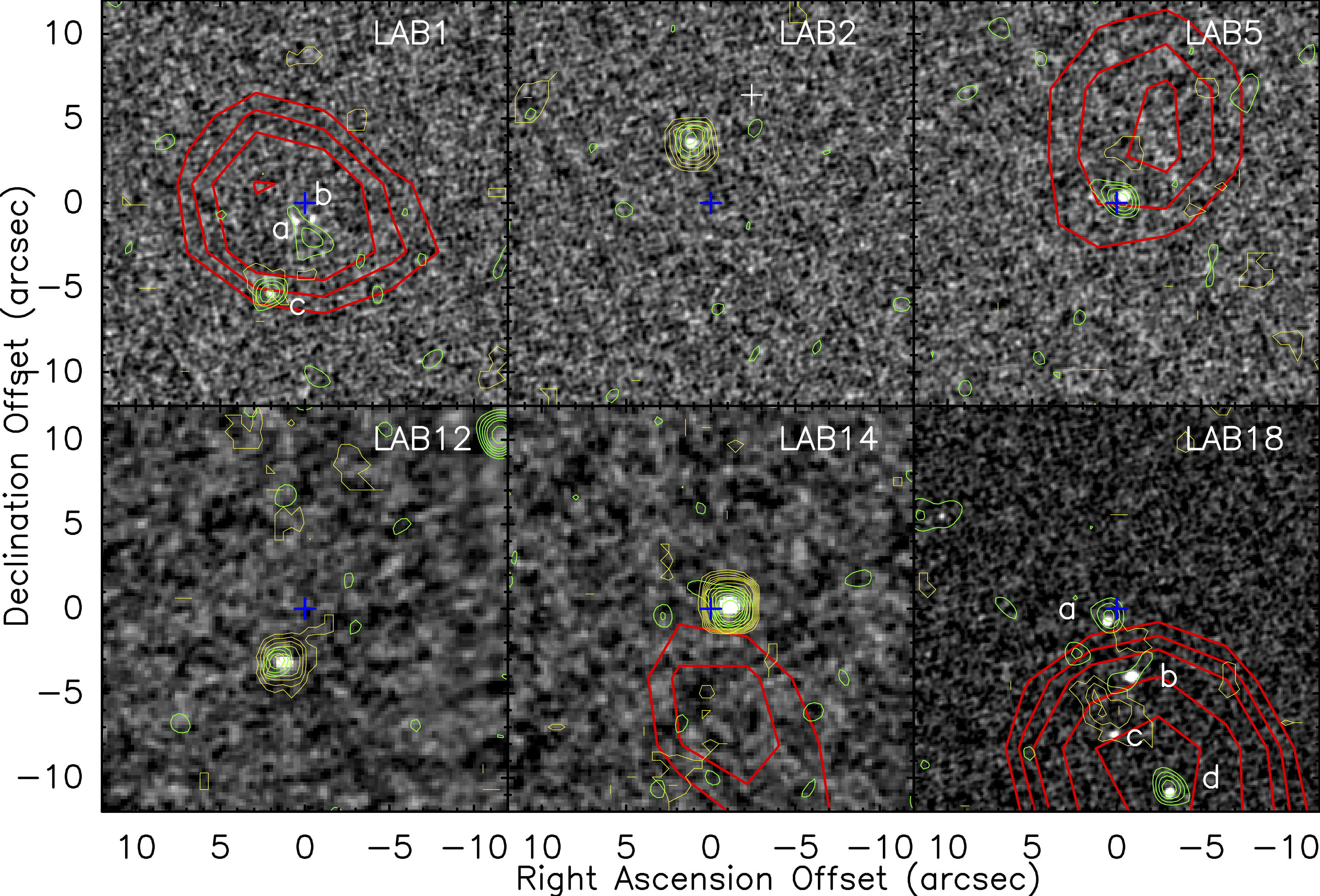}
\caption{SCUBA2 dust emission at 850~$\mu$m (red), VLA radio emission at 3~GHz
	(green) and $Chandra$ X-ray emission (yellow) contours are overlaid on
	the continuum images (LAB12 and LAB14 at 1.14mm and others at
	850~$\mu$m) of the LABs observed with ALMA. The offsets are relative to
	the centers of the LABs. The contour levels are the same as in
	Figure~3. ALMA dust continuum sources are labelled with
	letters in white when more than one dust core is detected in a single LAB.
	The white cross in the panel of LAB2 denotes the location of dust
	continuum marginally detected at 870~$\mu$m by Alexander et al. (2016)
	}\label{scuba_alma}	
\end{figure*}

Alternately, AGNs may also be responsible for heating the gas. Deep
$Chandra$ X-ray observations in the SSA 22 field (Geach et al.  2009; Lehmer et
al. 2009) cover 29 LABs. Indeed, 6 out of 29 LABs have X-ray counterparts, implying a
significant fraction of AGNs in LABs. Among these 6 X-ray detected LABs, LAB3
shows no counterparts at radio and submm wavelengths (for details see
$\S$~\ref{individual} and Figure~\ref{scuba_radio}) and LAB18 is not clearly
associated with any submm/radio counterparts, given the pointing accuracy of
the $Chandra$ observations (for details see $\S$~\ref{individual} and
Figures~\ref{scuba_radio} and \ref{scuba_alma}). The remaining four LABs, LAB2,
LAB12 and LAB14 identified in Geach et al. (2009), as well as LAB1 with weak
X-ray emission (see Figure~\ref{scuba_alma}), are detected at radio and submm
wavelengths.

\begin{figure*}[t]
\centering
\includegraphics[angle=0,width=1.0\textwidth]{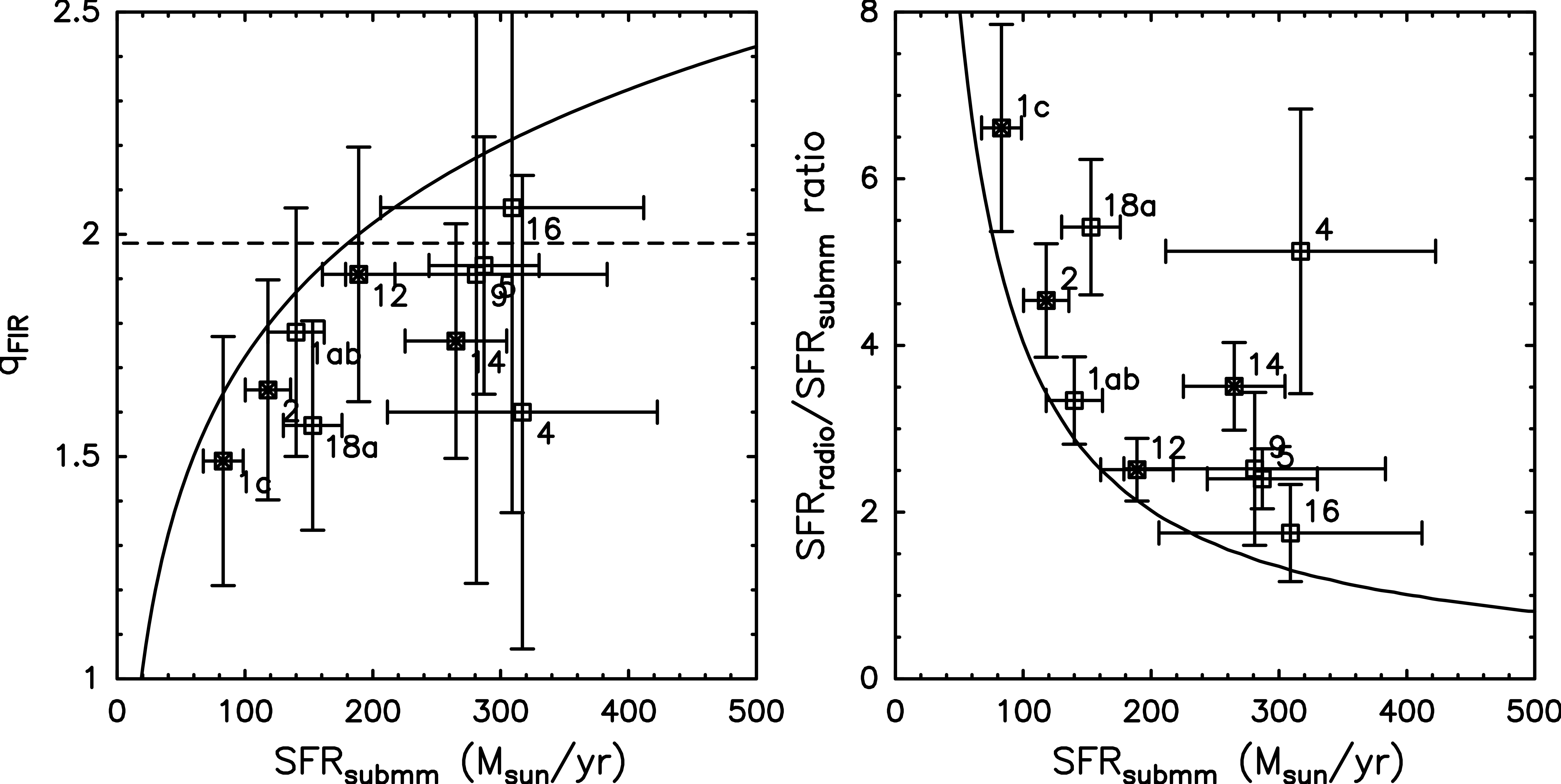}
	\caption{The parametrisation of the FRC, q$_{\rm FIR}$,
	({\bf left}) and SFR$_{\rm radio}$/SFR$_{\rm submm}$ ratios ({\bf
	right}) against SFR$_{\rm submm}$ for the LABs detected at dust and
	radio wavelengths. The sources detected in X-rays are shown with cross
	symbols inside the open squares. The solid lines present the detection
	limit, corresponding to a radio flux of 6.3~$\mu$Jy, constrained by the
	VLA observations. The dashed line in the left panel denotes the q$_{\rm
	FIR}$ value at $z$\,=\,3.1 predicted by the redshift evolution of
	q$_{\rm FIR}$ in Magnelli et al. (2015).
	}\label{ratio_sfr} 
\end{figure*}

The far-infrared (FIR) and radio luminosities of star forming
galaxies are tightly related via an empirical relationship, the FIR/radio
correlation (FRC; e.g Helou et al. 1988; Yun et al. 2001; Magnelli et al.
2015). The radio excess in the systems with AGNs drives them to deviate from
this tight relationship.  Therefore, following the method described in Magnelli
et al. (2015), we use the parametrisation of the FRC, q$_{\rm FIR}$, to study
the radio excess in our sample. The q$_{\rm FIR}$ parameter is defined as
\begin{equation}
	q_{\rm FIR}\,=\,{\rm log}(\frac{L_{\rm FIR}[{\rm W}]}{3.75\times10^{12}})-{\rm log}(L_{\rm 1.4GHz}[{\rm W\,Hz^{-1}}])
\end{equation}
(e.g. Helou et al. 1988; Yun et al. 2001; Magnelli et al. 2015), where $L_{\rm
FIR}$ is the integrated FIR luminosity from rest-frame 42 to 122 $\mu$m and
$L_{\rm IR}$=1.91$\times$$L_{\rm FIR}$ (Magnelli et al. 2015), and $L_{\rm
1.4GHz}$ is the rest-frame 1.4 GHz radio luminosity and is calculated
as in $\S$~\ref{sfr}. The results are shown in the left panel of
Figure~\ref{ratio_sfr}. We also compare our results to the q$_{\rm FIR}$ value
at $z$\,=\,3.1 predicted by the redshift evolution of q$_{\rm FIR}$ in Magnelli
et al. (2015). The detection limit set by the sensitivity of VLA observations
is shown as the solid line in Figure~\ref{ratio_sfr}. It is clear that all
X-ray detected LABs have q$_{\rm FIR}$ below the predicted value.  For the
sources detected at both submm and radio wavelengths, 8 out of 10 LABs have
q$_{\rm FIR}$ lower than the predicted value for the star forming galaxies,
suggesting a radio excess in LABs is common.  In the right panel of
Figure~\ref{ratio_sfr}, we plot SFR$_{\rm radio}$/SFR$_{\rm submm}$ ratios
against SFR$_{\rm submm}$. It is clear that X-ray detected LABs have SFR$_{\rm
radio}$/SFR$_{\rm submm}$ ratios above 2.5. For the sources detected at both
submm and radio wavelengths, 8 out of 10 LABs have ratios larger than 2.5,
suggesting a radio excess in LABs is common. Previously taken
radio/FIR data from local AGNs (Rush et al. 1993,1996) show that Seyfert
galaxies have a 3 times stronger 6cm radio continuum than predicted by star
formation alone, indicating that the existence of AGNs in submm/radio detected
LABs might be common. This is consistent with the finding that 10 out of 12
LAEs with $L_{\rm Ly\alpha}$$>$10$^{43}$\,ergs~s$^{-1}$ are detected in X-rays
with {\it Chandra} (Civano et al. 2016; Sorbral et al.  2017). However, it is
very difficult to discriminate between predominately SF or AGNs powered LABs.
At least for the submm/radio-detected sources with SFRs larger than
100~\msol/yr, it is very likely that SF is an important heating mechanism, and
among most of them AGNs may also play an important role in powering the \ly\,
emission.

Note that the upper SFR limits of the submm-undetected sources by the SCUBA-2
measurements are about 300\msol/yr, which is usually much higher than the SFR
determined by the \ly\, luminosity. Therefore, it is difficult to draw a strong
conclusion on the powering sources of these LABs with the current data, due
to the large variance in the \ly\, escape fraction (see
$\S$~\ref{lya_fraction}).  Future deep observations are required to settle this
problem.

\subsubsection{Locations of radio and X-ray counterparts within LABs and their
implications to heating mechanisms} 
Due to the limited angular resolution and positional uncertainties of the
SCUBA-2 observations, it is impossible to pinpoint the accurate positions of
the SMGs in the LABs with SCUBA-2 data alone. With the high angular resolution
VLA images, the accurate positions of radio counterparts and their associated
SMGs are now well determined. Among the sources with radio counterparts
detected above 4$\sigma$, four LABs, LAB4, LAB5, LAB9 and LAB16, have no X-ray
counterparts. Except for LAB9 the radio counterparts of the remaining three
sources, LAB4, LAB5, and LAB16, are located in the center of the LABs. This may
provide evidence in support of SF as the main powering source in these LABs,
and is consistent with the SF-based model (e.g., Zheng et al.  2011) for LAB
formation and recent observations by Matsuda et al. (2012) who find that
stacking of fainter LAEs in the SSA 22 field shows the extended and faint \ly\,
emission surrounding the bright sources at the center. However, the existence
of AGNs in these LABs can not be excluded. Especially, LAB4 has a high SFR$_{\rm
radio}$/SFR$_{\rm submm}$ ratio of 5.1\ppm1.7, strongly supporting that an AGN
may reside there. Thus, these sources without X-ray emission may also host
AGNs, but still deeply embedded in their host galaxies where X-ray emission
might be efficiently absorbed. This is supported by the fact that
these four LABs contain the high massive dust masses among all LABs.

X-ray  emission has been detected in 6 of the 29 LABs of the \ssa\, field.
This includes LAB1, not discussed in Geach et al. (2009),
suggesting the existence of AGNs in these systems. None of the X-ray sources is
located at the LAB's center. This positional inconsistency
between \ly\, and the embedded AGNs has already previously been found in a few LABs
(Prescott et al. 2012b; Yang et al. 2014).  In comparison with the X-ray
detected LABs, the LABs with the radio counterparts in the centers usually
contain more material and they are located in the centers of the LABs. It is
possible that their X-ray emission, if present, is largely absorbed by neutral
hydrogen, helium and presumably recently synthesized heavy elements. 
Alternatively, a recent simulation (Geach et al.
2016) shows that observed \ly\, surface brightness and morphology is highly
orientation dependent. This may lead to positional inconsistencies
between X-ray sources and the centers of LABs. However, this simulation can not
explain why most radio counterparts without X-ray emission locate at the
centers of LABs.

\subsection{\ly\, escape fraction at $z$=3.1}\label{lya_fraction}
\begin{figure*}[t]
\centering
\includegraphics[angle=0,width=1.0\textwidth]{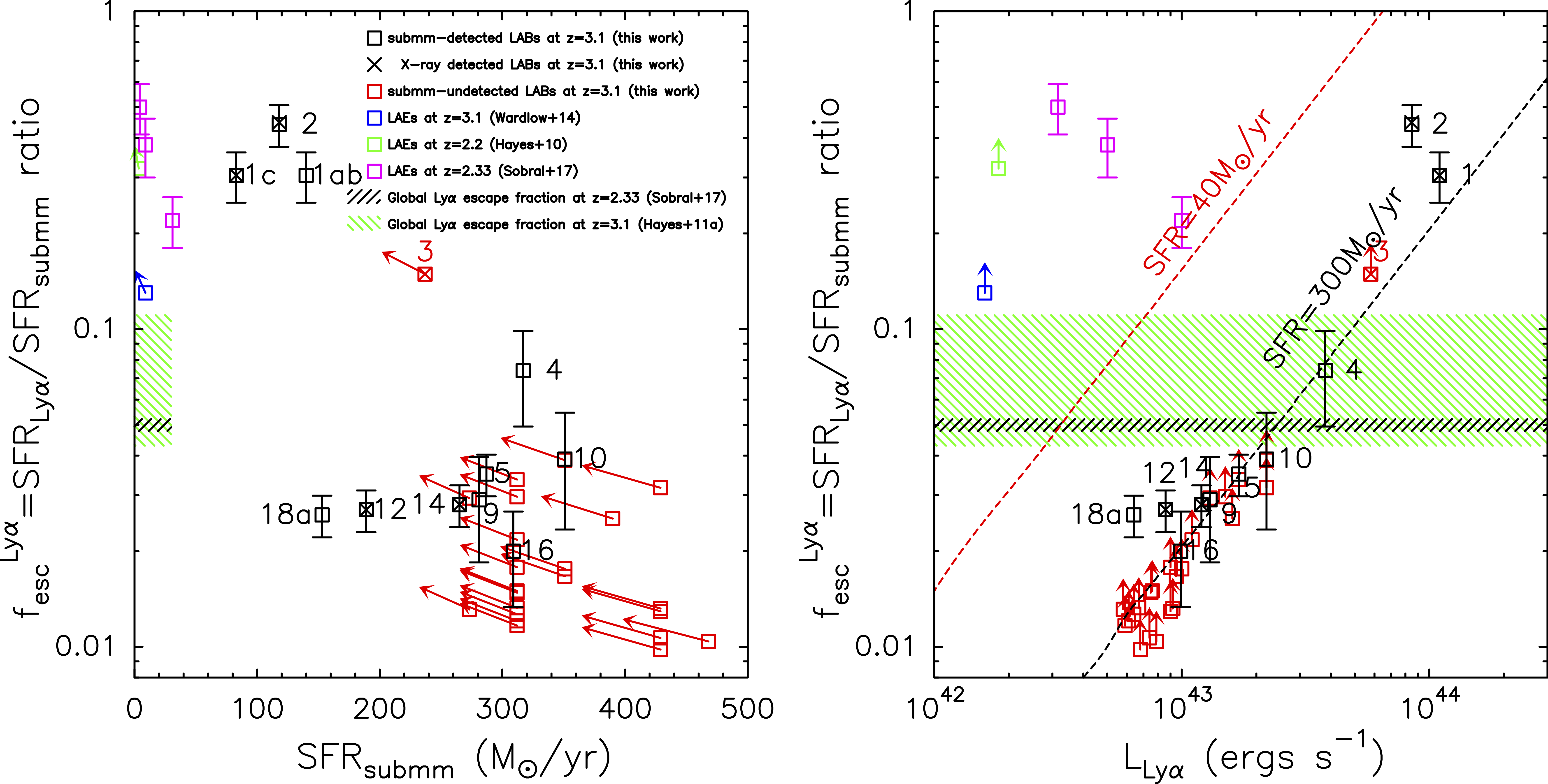}
\caption{Plots of escape fraction, $f_{\rm esc}^{\rm Ly\alpha}$, against star
	formation rate ({\bf left}) and \ly\, luminosity ({\bf right}).
	Submm-detected and submm-undetected LABs are marked by black and red
	squares, respectively.  X-ray detected LABs are indicated by crosses
	inside the squares. Lower limits of $f_{\rm esc}^{\rm Ly\alpha}$ for
	LAEs at $z$\,=\,2.2 and 3.1 are presented in green (Hayes et al. 2010)
	and blue (Wardlow et al. 2014), respectively. The fractions for LAEs
	at $z$\,=\,2.23 in Sobral et al. (2017) are shown in magenta. The
	global \ly\, escape fractions at $z$\,=\,2.33 (Sobral et al. 2017) and
	$z$\,=\,3.1 (Hayes et al. 2011a) are denoted by black and green hashed
	regions, respectively, where their typical SFRs are less
	than 20~\msol/yr. The dashed line in the right panel denotes the lower
	limit of $f_{\rm esc}^{\rm Ly\alpha}$ constrained by the SCUBA-2
	detection threshold of 300 \msol/yr. Some detected sources above this
	line are due to better sensitivities of the ALMA measurements. Some
	undetected LABs, shown as red open squares at 3$\sigma$,
	well below the line are limited by their poor sensitivities outside the
	central field of the SCUBA-2 observations.}\label{ratio}
\end{figure*}

It is very interesting to know the \ly\, escape fraction, $f_{\rm esc}^{\rm
Ly\alpha}$, in LABs, a special class of LAEs, and to understand it in different
environments. Traditionally, the \ly\, escape fraction at $z$$\ge$2.3 is mainly
derived from the \ly\, and UV/H$\alpha$ luminosity function (e.g., Hayes et
al.  2011a; Sobral et al.  2017). Dust emission is directly related to SF
activity in galaxies and has been widely used as a good SFR estimator
(Kennicutt 1998; Kennicutt \& Evans 2012). Therefore, using SCUBA-2/ALMA data,
the \ly\, escape fraction can also be calculated from the ratio of \ly\, to
dust continuum derived SFRs. In Figure~\ref{ratio}, we plot $f_{\rm esc}^{\rm
Ly\alpha}$ against the SFRs and \ly\, luminosities of the LABs.
The \ly\, escape fraction ranges from $~$2\% to 45\% among the submm-detected
LABs, with a median value of about 3\%. Except for 11 submm-detected sources,
the remaining LABs not detected at submm wavelengths can also provide
important constraints on $f_{\rm esc}^{\rm Ly\alpha}$. Lower limits of $f_{\rm
esc}^{\rm Ly\alpha}$ can be estimated by using the upper limits of SCUBA-2
measurements and these values are presented in Figure~\ref{ratio}.
We also stack the submm-undetected LABs, but fail to detect the
dust emission at $>$3$\sigma$, leading to a lower limit of $f_{\rm esc}^{\rm
Ly\alpha}$ to be 11\%. This suggests that the majority of the sample have
higher $f_{\rm esc}^{\rm Ly\alpha}$ in comparison with the global \ly\,
escape fraction in the previous studies (Hayes et al. 2011a; Sobral et al.
2017), where the global escape fraction refers to the value for
all galaxies in a field, and is usually determined from the integral of the
\ly\, and H$\alpha$ luminosity functions. Indeed, the fractions in the LAEs
are usually much higher than the global values, as the LAEs are a biased
sample, selected by their strong \ly\, emission. Hayes et al. (2010) reported
$f_{\rm esc}^{\rm Ly\alpha}$ $>$ 32\% for LAEs at $z$=2.2, Sobral et al. (2017)
shows that $f_{\rm esc}^{\rm Ly\alpha}$ is around 37\% for LAEs at $z$=2.23.
Wardlow et al. (2014) use far-infrared data instead of optical data and
obtain a lower limit of $f_{\rm esc}^{\rm Ly\alpha}$ to be 10 to 20\% for LAEs
at $z$=3.1, which is consistent with our stacking results
for the majority of the LABs.

We notice that most submm-detected LABs have \ly\, escape
fractions much lower than the fractions of LAEs in other studies (e.g.
Hayes et al. 2010; Wardlow et al. 2014; Sobral et al 2017), as shown in
Figure~\ref{ratio}. We find a big difference between these LABs and those of
Hayes et al.  (2010), Wardlow et al. (2014) and Sobral et al (2017) that the
SFRs of the submm-detected LABs are much higher than those in the latter three
samples. The galaxies with high SFRs are most likely massive galaxies with a
large amount of dust. We suspect that the low \ly\, escape fractions in the
submm-detected LABs are probably due to high dust attenuation as their SFRs are
about one magnitude higher than in the samples of other studies.  Another
reason may be related to the large-scale environment around the LABs.  All
sources are located in the dense region of SSA~22, which contains a large
amount of circumgalactic medium. In such an environment, \ly\, photons
experience numerous resonance scatterings, and can also be easily absorbed by
surrounding neutral hydrogen atoms, making the \ly\, emission
more extended and leading to a lower escape fraction. This is consistent with
the recent study of Shimakawa et al. (2017) who find that \ly\, escape
fractions in high-density regions are lower than in low-density regions.  We
also note that three LABs, LAB1, LAB2 and LAB3, have similar \ly\, escape
fractions as the LAEs studied by Hayes et al. (2010), Wardlow et al. (2014) and
Sobral et al (2017). All these sources have X-ray emission. Their high
fractions may be partially due to the AGNs in the host galaxies. The AGNs will
produce additional photons and feedback to possibly remove the material
surrounding the centers and then enhance the \ly\, emission by increasing the
escaping photons from the central regions. We also need to
mention that our SCUBA-2 data provide superior measures of SFRs from
extinction-free dust emission, in comparison with typically used optical data
that suffer large uncertainties due to the extinction correction. The actual escape
fractions will decrease if considering an extra extinction related to the
H{\scriptsize II} regions for H$\alpha$ (Wuyts et al. 2013; Reddy et al. 2015,
2016; An et al. 2017). To better understand the physical origin of the low
\ly\, escape fraction in the LABs, deep submm observations will be useful.

\section{Conclusions}

We have presented submm data from JCMT/SCUBA-2 observations together with ALMA
data towards 35 LABs identified in the \ssa\, field by Matsuda et al. (2004)
and deep radio images with the VLA towards 29 out of the 35 LABs. We also
discuss the deep X-ray data from the same region. Our conclusions are as
follows:

\begin{enumerate}
\item With the JCMT/SCUBA-2 and ALMA, 11 out of 35 LABs show dust emission.
	With the VLA, 9 out of 29 LABs are detected at radio wavelengths. For 5
		out of the 9 radio detected sources, spectroscopic data from
		the literature reveal redshifts around 3.1, confirming that the
		radio counterparts and the corresponding submm sources are
		associated with the LABs.

\item The detection of submm dust emission in LABs is more linked to the
	physical sizes of \ly\, emission than their \ly\, luminosities.

\item The 11 LABs detected at submm wavelengths have SFRs over 100~\msol/yr,
	favoring star formation as an important energy source
		for some LABs. Our results show that a radio excess is common
		in the submm/radio detected LABs and therefore AGNs may also
		play an important role to power the \ly\, emission.

\item Among the four radio detected LABs lacking X-ray emission, three sources
	are located in the center of their parent LABs. All X-ray sources are
		not located at the center of their associated LABs.
		The latter may be explained by absorption due to
		neutral hydrogen, helium and potentially newly formed heavy
		elements or by LAB morphologies, which are highly orientation
		dependent.
		
\item The \ly\, escape fraction ranges from $~$2\% to 45\% among
	the submm-detected LABs, with a median value of about 3\%.  Based on
		the stacked data, the submm-undetected LABs show a lower limit
		of 11\%, which is consistent with those of LAEs in previous
		studies. However, some submm-detected LABs have significantly
		lower \ly\, escape fraction. We suspect that this large
		variation in \ly\, escape fractions is due to the high dust
		attenuation supported by the large SFRs in our sample, the
		dense large-scale environment in \ssa\, as well
		as large uncertainties related to the extinction correction in
		optical data.
	
\end{enumerate}

\begin{acknowledgements}
We thank the anonymous referee for valuable comments that improved this
	manuscript.  Y.A. thanks Fangxia An for the comment about the effect of
	extinction correction on \ly\, escape fraction and Thomas Reiprich for
	the discussion about the contributions to the X-ray emission. Y.A. was
	supported by the ALMA Japan Research Grant of NAOJ Chile Observatory,
	NAOJ-ALMA-0165. Y.A. acknowledges partial support by NSFC grant
	11373007 and Youth Innovation Promotion Association CAS.  M.H.
	acknowledges the support of the Swedish Research Council,
	Vetenskapsr{\aa}det and the Swedish National Space Board (SNSB), and is
	Fellow of the Knut and Alice Wallenberg Foundation. N.K.H. is supported
	by the Science and Technology Facilities Council (grant number
	ST/K502029/1).

The James Clerk Maxwell Telescope has historically been operated by the Joint
Astronomy Centre on behalf of the Science and Technology Facilities Council of
the United Kingdom, the National Research Council of Canada and the Netherlands
Organisation for Scientific Research. Additional funds for the construction of
SCUBA-2 were provided by the Canada Foundation for Innovation.
The James Clerk Maxwell Telescope is operated by the East Asian Observatory on
behalf of The National Astronomical Observatory of Japan, Academia Sinica
Institute of Astronomy and Astrophysics, the Korea Astronomy and Space Science
Institute, the National Astronomical Observatories of China and the Chinese
Academy of Sciences (Grant No. XDB09000000), with additional funding support
from the Science and Technology Facilities Council of the United Kingdom and
participating universities in the United Kingdom and Canada.

This paper makes use of the following ALMA data: ADS/JAO.ALMA\#2013.1.00162.S
and ADS/JAO.ALMA\#2013.1.00704.S. ALMA is a partnership of ESO (representing its
member states), NSF (USA) and NINS (Japan), together with NRC (Canada), NSC and
ASIAA (Taiwan), and KASI (Republic of Korea), in cooperation with the Republic
of Chile. The Joint ALMA Observatory is operated by ESO, AUI/NRAO and NAOJ.

\end{acknowledgements}

\clearpage

\clearpage

\begin{center}
\begin{table*}[t]
\centering
\caption{Observation logs in \ssa}\label{table_log}
\begin{tabular}{ccccccccc}
\hline
	Telescope & Observation date & Project & RA   & Dec.    &  Freq. & Map size$^{a}$ &  Observing time & On-source time \\ 
		  &                  &      & (J2000) & (J2000) &  (GHz) & (arcmin) &  (hours) & (hours) \\ 	
\hline
	JCMT/SCUBA-2 & 2012.09 to 2013.12 &  MJLSC02 & 22:17:36.30 & +00:19:22.7 & 345 & 30 & ...  & 72  \\ 
	JCMT/SCUBA-2 & 2015.04 to 2015.06 &  M15AI91 & 22:17:31.70 & +00:17:50.0 & 345 & 15 & ...  & 19  \\ 
	VLA & 2015.04 to 2015.04 &  15A-120 & 22:17:28.00 & +00:17:50.0 & 2$-$4   & 15 & 6.7  & 4.2 \\
	VLA & 2016.05 to 2016.05 &  16A-310 & 22:17:43.00 & +00:10:40.0 & 2$-$4   & 15 & 2    & 1.6 \\
	VLA & 2016.05 to 2016.08 &  16A-310 & 22:17:32.00 & +00:15:00.0 & 2$-$4   & 15 & 32   & 26.1 \\
\hline
\end{tabular}
\begin{list}{}{}
\item{$^a$ The map size of the VLA observations is the Field of View at the central frequency of 3~GHz.}
\item{Note that the best sensitivities in the images are 0.75~mJy/beam for the combined SCUBA-2 observations
	and 1.5~$\mu$Jy/beam for the combined VLA observations, respectively.}
\end{list}
\end{table*}
\end{center}

\begin{longrotatetable}
\begin{deluxetable*}{ccccccccccccccccccc}
\centering
\tablecaption{JCMT/SCUBA-2 and VLA S-band observations towards LABs in \ssa\,\label{table_obs}}
\tabletypesize{\tiny}
\tablehead{
	& & & &  \multicolumn{4}{c}Previous (sub)mm measurements$^{b}$ & \multicolumn{6}{c}{VLA observations$^{c}$} &  \multicolumn{4}{c}{Spectroscopic observations$^{d}$} \\
	\cline{5-7} \cline{9-13}   \cline{15-18}  \\ 	
	Source & RA(J2000) & Dec(J2000) & S$_{{850\mu}m}$$^{a}$ & S$_{{850\mu}m}$ & S$_{{850\mu}m}$ & S$_{1.1mm}$ & & VLA ID & rms & offx & offy &  S$_{10cm}$ &  & offx & offy & zred & ref \\
	&           &            &  (mJy)  &  (mJy)  &  (mJy)  &  (mJy)  &        &      & ($\mu$Jy) & ($\arcsec$) & ($\arcsec$) &  ($\mu$Jy) & & ($\arcsec$) & ($\arcsec$) &  &  \\
	& & & &	Geach2005 & Hine2016 & Tamura2013 
}
\startdata
	SCUBA2-LAB1&   22:17:25.961&   +00:12:37.57&  {\bf 2.9\ppm0.8}    & {\bf 16.8\ppm2.9}  & {\bf 4.6\ppm1.1}  & 1.97\ppm0.74  &    \\
	ALMA-LAB1ab&   22:17:25.981 & +00:12:36.35 &  {\bf 1.08\ppm0.17}  &   &   &   &     & {\bf VLA-LAB1a}  &1.64/1.79& -0.4 & -1.1 &   7.3\ppm2.2 \\
	ALMA-LAB1c&   22:17:26.100 & +00:12:32.37 &  {\bf 0.64\ppm0.12}   &   &   &   &     & {\bf VLA-LAB1b}  &1.64/1.79&  0.2 & -0.1 &   8.6\ppm2.2& &   0.0 &  -0.1 &   3.0993+/-0.0004 &    1 \\
	ALMA-LAB1&      &   &  {\bf 1.72\ppm0.21} \\	
   SCUBA2-LAB2&   22:17:38.996 &   +00:13:27.51& -0.5\ppm0.8              & 3.3\ppm1.2  & 0.1\ppm1.1  & -1.89\ppm0.76  &     \\
	ALMA-LAB2&   22:17:39.079 &  +00:13:30.85 & {\bf 0.91\ppm0.10}    &   &   &   &     & {\bf VLA-LAB2}  &1.57/1.65&  0.0 &  0.0 &   8.4\ppm2.4& &   -0.5 &  -0.7 &     3.091+/-0.001&  3 \\
   SCUBA2-LAB3&   22:17:59.153&   +00:15:28.37&  0.2\ppm0.9               & -0.2\ppm1.5  & 0.1\ppm1.1  & -0.69\ppm0.73  &      & VLA-LAB3  &1.54/2.70& -1.9& -1.8&   7.5\ppm3.8 \\
	SCUBA2-LAB4&   22:17:25.126&   +00:22:10.21&  {\bf 2.4\ppm0.8}    & 0.9\ppm1.5  & 2.4\ppm1.1  & 0.11\ppm0.74  &     & {\bf VLA-LAB4a}  &1.48/2.72&  0.0&  0.1&  25.4\ppm4.1 \\
                                               &&& &   &   &   &      & VLA-LAB4b  &1.48/2.72&  2.3& -5.6&  98.9\ppm3.8 \\
	SCUBA2-LAB5&   22:17:11.681&   +00:16:43.95&  {\bf 2.9\ppm0.8}  & {\bf 5.2\ppm1.4}  & 1.9\ppm1.1  & 0.34\ppm0.74  &      \\
	ALMA-LAB5&   22:17:11.664 & +00:16:44.32 &  {\bf 2.21\ppm0.08}  &   &   &   &     & {\bf VLA-LAB5}  &1.29/1.93&  0.1 &  -0.4 &  10.8\ppm2.7 \\
   SCUBA2-LAB6&   22:16:51.428&   +00:25:02.39&  0.4\ppm1.0             & -0.5\ppm1.8  & 1.0\ppm1.1  & 0.07\ppm1.14  &     & &n.a.& \\
   SCUBA2-LAB7&   22:17:41.005&   +00:11:26.32& -1.0\ppm0.8             & 0.2\ppm1.6  & 1.2\ppm1.1  & -0.88\ppm0.74  &     & &1.52/1.81& \\
   SCUBA2-LAB8&   22:17:26.176&   +00:12:53.53& -0.6\ppm0.8             & 0.3\ppm5.3  & 2.6\ppm1.1  & 0.67\ppm0.74  &     & &1.65/1.78& \\
	SCUBA2-LAB9&   22:17:51.084&   +00:17:26.31&  {\bf 2.2\ppm0.8}  & 1.3\ppm5.3  & 2.2\ppm1.1  & 0.07\ppm0.74  &     & {\bf VLA-LAB9a}  &1.54/2.25& -2.0&  0.7&  11.1\ppm3.4 \\
					       &&&                      &   &   &    &   & VLA-LAB9b  &1.54/2.25&  2.6&  0.0&   7.0\ppm3.4 \\
	SCUBA2-LAB10&   22:18:02.250&   +00:25:55.77&  {\bf 2.7\ppm1.1} & {\bf 6.1\ppm1.4}  & 3.2\ppm1.1  & 1.20\ppm0.84  &     & &n.a.& \\
	SCUBA2-LAB11&   22:17:20.325&   +00:17:32.05&  {\bf 2.3\ppm0.7} & -0.4\ppm5.3  & 2.5\ppm1.1  & 0.61\ppm0.73  &    &  VLA-LAB11  &1.46/1.78& -1.8&  2.9&   5.5\ppm2.7 \\
   SCUBA2-LAB12&   22:17:31.907&   +00:16:58.77&  0.7\ppm0.8            & 3.2\ppm1.6  & 0.8\ppm1.1  & 0.30\ppm0.74  &     \\
	ALMA-LAB12 & 22:17:32.01  & +00:16:55.4 & {\bf 0.63\ppm0.03}      &   &   &   &    & {\bf VLA-LAB12}  & 1.61/1.70&  0.2 & 0.1 &   7.4\ppm2.4& &   -0.2 &  0.1 &   3.0909+/-0.0004&    2 \\
   SCUBA2-LAB13&   22:18:07.972&   +00:16:46.77&  0.1\ppm0.9            &   & 1.2\ppm1.1  & -0.72\ppm0.73    &  & &1.32/3.95& \\
	SCUBA2-LAB14&   22:17:35.908&   +00:15:58.79&  {\bf 2.0\ppm0.7} & {\bf 4.9\ppm1.3}  & 2.0\ppm1.1  & 2.43\ppm0.76    &   \\
	ALMA-LAB14 &  22:17:35.83 & +00:15:59.0 & {\bf 1.45\ppm0.09}      &   &   &   &    &  {\bf VLA-LAB14}  &1.51/1.54& 0.1 & -0.4 &  14.5\ppm2.1& &  1.0 &   0.0 &             3.094&   4 \\
   SCUBA2-LAB15&   22:18:08.317&   +00:10:21.78&  1.3\ppm1.1            &   & 2.4\ppm1.1  & -0.27\ppm0.74  &   &  &1.51/4.33& \\
	SCUBA2-LAB16&   22:17:24.845&   +00:11:16.77&  {\bf 2.4\ppm0.8} & 2.2\ppm5.3  & 3.4\ppm1.1  & 0.34\ppm0.74  &    & {\bf VLA-LAB16}  &1.44/1.77&  0.2&  0.6&   8.5\ppm2.7& &   0.8&   0.7&   3.0689+/-0.0002&   2 \\
   SCUBA2-LAB17&   22:18:36.533&   +00:07:19.88& -1.5\ppm1.9            &   &   & 1.41\ppm1.19  &   & &n.a.& \\
	SCUBA2-LAB18&   22:17:28.998 &   +00:07:51.16&  {\bf 5.4\ppm0.9}& {\bf 11.0\ppm1.5}  & {\bf 5.2\ppm1.1}  & 2.33\ppm0.73  &    \\
	ALMA-LAB18a &   22:17:29.032 &  +00:07:50.26 &  {\bf 1.18\ppm0.08}&   &   &   &     & {\bf VLA-LAB18a}  &1.58/2.80&  -0.2& 0.1&  13.0\ppm4.2 \\
	ALMA-LAB18b &   22:17:28.936 & +00:07:46.92 &  {\bf 2.73\ppm0.10}&   &   &   &      &  VLA-LAB18b  &1.58/2.80& -0.1& 0.0 &   9.1\ppm4.0 \\
	ALMA-LAB18c &   22:17:29.017 & +00:07:43.44 &  {\bf 1.14\ppm0.15}&   &   &   &     & &1.58/2.80 \\
	ALMA-LAB18d &   22:17:28.781 & +00:07:40.07 &  {\bf 4.42\ppm0.30}&   &   &   &     & VLA-LAB18d  &1.58/2.80& 0.2 & 0.4 &  16.4\ppm4.3 \\
	ALMA-LAB18 &     &   &  {\bf 9.47\ppm0.36}                       &   &   &   &    \\	
   SCUBA2-LAB19&   22:17:19.569&   +00:18:46.38& -2.1\ppm0.7             & -8.6\ppm5.3  & -0.4\ppm1.1  & -0.81\ppm0.74  &    & &1.56/2.09& \\
   SCUBA2-LAB20&   22:17:35.307&   +00:12:48.31& -0.4\ppm0.8             & 0.4\ppm1.5  & 0.2\ppm1.1  & -0.80\ppm0.75  &    & &1.60/1.68& \\
   SCUBA2-LAB21&   22:18:17.324&   +00:12:08.66&  1.1\ppm1.2             &   & 0.9\ppm1.1  & -1.37\ppm0.75  &   &  &1.63/6.63& \\
   SCUBA2-LAB22&   22:17:34.982&   +00:23:35.09&  1.5\ppm0.8             &   & 1.3\ppm1.1  & 1.04\ppm0.74  &   &  &1.31/3.45& \\
   SCUBA2-LAB23&   22:18:07.950&   +00:23:16.62& -0.1\ppm1.1             &   & 1.0\ppm1.1  & -1.55\ppm0.80  &    & &n.a.& \\
   SCUBA2-LAB24&   22:18:00.905&   +00:14:40.10& -0.9\ppm0.8             &   & -0.6\ppm1.1  & 0.03\ppm0.72  &    & &1.64/2.93& \\
   SCUBA2-LAB25&   22:17:22.590&   +00:15:50.86& -2.1\ppm0.8             & 1.4\ppm5.3  & -1.5\ppm1.1  & 0.01\ppm0.73  &   &  &1.43/1.57& \\
   SCUBA2-LAB26&   22:17:50.424&   +00:17:33.37&  0.6\ppm0.8             & -2.7\ppm5.3  & 1.1\ppm1.1  & -0.90\ppm0.74  &    & &1.50/2.17& \\
   SCUBA2-LAB27&   22:17:06.974&   +00:21:30.15&  1.7\ppm0.8             & 0.5\ppm1.6  & 2.1\ppm1.1  & 0.18\ppm0.77  &    & &1.29/3.70& \\
   SCUBA2-LAB28&   22:17:59.210&   +00:22:53.96& -0.1\ppm0.9             &   & -0.6\ppm1.1  & -0.99\ppm0.76  &   &  &1.41/5.83& \\
   SCUBA2-LAB29&   22:16:53.869&   +00:23:00.39&  0.3\ppm1.1             &   & 0.7\ppm1.1  & -2.54\ppm0.91  &   &  &n.a.& \\
   SCUBA2-LAB30&   22:17:32.454&   +00:11:33.36&  0.6\ppm0.9             & 3.3\ppm1.3  & 1.9\ppm1.1  & 0.65\ppm0.74  &    & VLA-LAB30a  &1.72/1.91&  4.6&  1.6&  24.9\ppm2.4 \\
					       &&&&   &   &   &     & VLA-LAB30b  &1.72/1.91&  1.1&  0.7&   4.3\ppm2.4 \\
                                               &&&&   &   &   &     & VLA-LAB30c  &1.72/1.91&  3.4& -2.5&  20.0\ppm2.6 \\
                                               &&&&   &   &   &     & VLA-LAB30d  &1.72/1.91& -4.8& -3.7&  46.9\ppm2.6 \\
   SCUBA2-LAB31&   22:17:38.945&   +00:11:01.87& -0.7\ppm0.8            & -3.7\ppm5.3  & 0.0\ppm1.1  & -1.44\ppm0.74  &   &  &1.48/1.78& \\
   SCUBA2-LAB32&   22:17:23.874&   +00:21:55.46&  0.6\ppm0.7            & 1.8\ppm1.4  & 0.9\ppm1.1  & -0.16\ppm0.74  &   &  &1.50/2.75& \\
   SCUBA2-LAB33&   22:18:12.553&   +00:14:32.67& -0.6\ppm1.1            & 1.6\ppm1.5  & 0.7\ppm1.1  & 0.04\ppm0.73  &   &  &1.41/4.92& \\
   SCUBA2-LAB34&   22:16:58.365&   +00:24:29.08& -0.7\ppm1.1            &   & 0.4\ppm1.1  & 1.01\ppm0.93  &   &  &n.a.& \\
   SCUBA2-LAB35&   22:17:24.837&   +00:17:16.92& -0.3\ppm0.8            & 1.2\ppm5.3  & 1.0\ppm1.1  & -0.74\ppm0.73  &   &  &1.58/1.77& \\
\enddata
\begin{list}{}{}
\item{$^{\mathrm{a}}$ The fluxes are measured at 850$\mu$$m$ by JCMT/ALMA
	except for LAB12 and LAB14. The latter two sources were observed at
	1.14mm with ALMA (Umehata et al. 2015). The sources detected at submm wavelengths are
	highlighted in bold font. 
	$^{\mathrm{b}}$ Previous (sub)millimeter measurements
	from three deep surveys. The first two were carried out at 850$\mu$$m$
	by SCUBA (Geach et al. 2005) and SCUBA2 (Hine et al. 2016), respectively. The third
	one was taken at 1.1~mm by AzTEC/ASTE (Tamura et al. 2013). Sources
	detected with signal-to-noise ratios at $\ge$3.5$\sigma$ are shown in
	bold font.
        $^{\mathrm{c}}$ In Column~9, the given two values are for the noise levels
	before and after primary beam correction. The offsets of radio
	counterparts relative to the LAB centers or ALMA sources are given
	in Column~7 \& 8. The counterparts associated with the LABs are
	highlighted in bold font.
	$^{\mathrm{d}}$ Only the targets with offsets (Column~13 \& 14) less than
	1 arcsec relative to the radio sources are considered to be associated
	with the radio sources. References: 1. Umehata et al. (2017b); 2. Kubo
	et al. (2015); 3. Steidel et al. (2003); 4. Yamada et al. (2012)}
\end{list}
\end{deluxetable*}
\end{longrotatetable}

\begin{center}
\begin{table*}[t]
\centering
	\caption{Derived star formation rates towards submm/radio-detected LABs in \ssa}\label{table_sfr}
\begin{tabular}{ccccccc}
\hline
	Source & SFR$_{{\rm 850\mu}m}$$^{a}$ & Radio comp & SFR$_{\rm
	10cm}$$^{a}$ & SFR$_{\rm 10cm}$/SFR$_{{\rm 850\mu}m}$  & SFR$_{\rm
	Ly\alpha}$$^{a}$ & SFR$_{\rm Ly\alpha}$/SFR$_{{\rm 850\mu}m}$  \\
		& (\msol/yr) & &  &  (\msol/yr) &  (\msol/yr) &  (\%) \\
\hline
SCUBA2-LAB1& 376$^{+332}_{-99}$ & VLA-LAB1 & 1017\ppm198 & 2.7   & 68 &   18.1 \\
ALMA-LAB1ab& 140$^{+124}_{-37}$& VLA-LAB1a & 468\ppm140 & 3.3 \\
ALMA-LAB1c&  83$^{+73}_{-22}$& VLA-LAB1b & 549\ppm140 & 4.6 \\
ALMA-LAB1& 223$^{+197}_{-59}$ & VLA-LAB1 & 1017\ppm198 & 4.0    & 68 & 30.5 \\
SCUBA2-LAB2&$<$195 & VLA-LAB2 & 536\ppm156 & $>$2.7             &  52   & $>$26.5 \\
ALMA-LAB2& 118$^{+104}_{-31}$& VLA-LAB2 & 536\ppm156 & 4.5      & 52  & 44.1 \\
SCUBA2-LAB3$^{b}$&$<$237 & ... & $<$482 & ...                         &  36  & $>$14.9 \\
SCUBA2-LAB4& 317$^{+280}_{-84}$ & VLA-LAB4a &1625\ppm264 & 5.1  &  23 &   7.4 \\
                    && VLA-LAB4b &6342\ppm242 \\
SCUBA2-LAB5& 382$^{+338}_{-101}$ & VLA-LAB5 & 690\ppm175 & 1.8  &  10 &    2.6 \\
ALMA-LAB5 & 287$^{+254}_{-76}$& VLA-LAB5 & 690\ppm175 & 2.4     &  10 &    3.5 \\
SCUBA2-LAB9& 281$^{+248}_{-74}$ & VLA-LAB9a & 709\ppm215 & 2.5  &  8.0 &   2.9 \\
                    && VLA-LAB9b & 450\ppm218 \\
	SCUBA2-LAB10& 349$^{+308}_{-92}$ & ...  & ... &  ...    & 14 &    3.9 \\
SCUBA2-LAB11& 302$^{+266}_{-80}$ & VLA-LAB11 & 352\ppm172 & 1.2  &  5.6 &    1.8 \\
SCUBA2-LAB12&$<$214 & VLA-LAB12 & 474\ppm156 & $>$2.2            &  5.3 & $>$2.5 \\
	ALMA-LAB12& 189$^{+76}_{-19}$& VLA-LAB12 & 474\ppm156 & 2.5 & 5.3 & 2.7 \\
SCUBA2-LAB14& 265$^{+234}_{-70}$ & VLA-LAB14 & 930\ppm137 & 3.5     &  7.4 &    2.8 \\
	ALMA-LAB14 & 389$^{+227}_{-63}$& VLA-LAB14 & 930\ppm137 & 2.4 & 7.4 & 1.9 \\
SCUBA2-LAB16& 309$^{+273}_{-82}$ & VLA-LAB16 & 541\ppm170 & 1.7       &   6.1 &    2.0 \\
SCUBA2-LAB18& 696$^{+615}_{-184}$ & VLA-LAB18abd & 2464\ppm462 & 3.5  \\ 
ALMA-LAB18a& 153$^{+135}_{-40}$& VLA-LAB18a & 830\ppm270 & 5.4       & 4.0 & 2.6 \\
ALMA-LAB18b& 355$^{+313}_{-94}$& VLA-LAB18b & 586\ppm254 & 1.7 \\
ALMA-LAB18d& 575$^{+508}_{-152}$& VLA-LAB18d &1048\ppm275 & 1.8 \\

\hline
\end{tabular}
\begin{list}{}{}
\item{$^{\mathrm{a}}$ The star formation rates (SFRs) are determined from
	measured submm and radio fluxes under some assumptions (for details,
	see the text in $\S~\ref{sfr}$. 2$\sigma$ upper limits are given
	for submm-undetected LABs. $^{\mathrm{b}}$ Note that the LAB, not
	detected at dust and radio wavelengths but with X-ray emission, is also
	presented in this Table.}
\end{list}
\end{table*}
\end{center}

\end{document}